\documentclass[twocolumn]{aa}

\usepackage{graphicx}
\usepackage{amsmath,amsfonts,amssymb}
\usepackage{txfonts}
\usepackage{color}
\usepackage{natbib}
\usepackage{float}
\usepackage{subfigure} %added, uf
\usepackage{dblfloatfix}
\usepackage{afterpage}
\usepackage{ifthen}
\usepackage[morefloats=12]{morefloats}
\usepackage{placeins}
\usepackage{multicol}
\bibpunct{(}{)}{;}{a}{}{,}
\usepackage[switch]{lineno}
\definecolor{linkcolor}{rgb}{0.6,0,0}
\definecolor{citecolor}{rgb}{0,0,0.75}
\definecolor{urlcolor}{rgb}{0.12,0.46,0.7}
\usepackage[breaklinks, colorlinks, urlcolor=urlcolor,
    linkcolor=linkcolor,citecolor=citecolor,pdfencoding=auto]{hyperref}
\hypersetup{linktocpage}

\def\setsymbol#1#2{\expandafter\def\csname #1\endcsname{#2}}
\def\getsymbol#1{\csname #1\endcsname}

\def\Planck{\textit{Planck}}

\newbox\tablebox    \newdimen\tablewidth
\def\leaderfil{\leaders\hbox to 5pt{\hss.\hss}\hfil}
\def\endPlancktablewide{\tablewidth=\textwidth 
    $$\hss\copy\tablebox\hss$$
    \vskip-\lastskip\vskip -2pt}
\def\tablenote#1 #2\par{\begingroup \parindent=0.8em
    \abovedisplayshortskip=0pt\belowdisplayshortskip=0pt
    \noindent
    $$\hss\vbox{\hsize\tablewidth \hangindent=\parindent \hangafter=1 \noindent
    \hbox to \parindent{$^#1$\hss}\strut#2\strut\par}\hss$$
    \endgroup}
\def\doubleline{\vskip 3pt\hrule \vskip 1.5pt \hrule \vskip 5pt}

\def\L2{\ifmmode L_2\else $L_2$\fi}

\def\DeltaT{\ifmmode \Delta T\else $\Delta T$\fi}
\def\deltat{\ifmmode \Delta t\else $\Delta t$\fi}
\def\fknee{\ifmmode f_{\rm knee}\else $f_{\rm knee}$\fi}
\def\Fmax{\ifmmode F_{\rm max}\else $F_{\rm max}$\fi}
\def\solar{\ifmmode{\rm M}_{\mathord\odot}\else${\rm M}_{\mathord\odot}$\fi}
\def\Msolar{\ifmmode{\rm M}_{\mathord\odot}\else${\rm M}_{\mathord\odot}$\fi}
\def\Lsolar{\ifmmode{\rm L}_{\mathord\odot}\else${\rm L}_{\mathord\odot}$\fi}
\def\inv{\ifmmode^{-1}\else$^{-1}$\fi}
\def\mo{\ifmmode^{-1}\else$^{-1}$\fi}
\def\sup#1{\ifmmode ^{\rm #1}\else $^{\rm #1}$\fi}
\def\expo#1{\ifmmode \times 10^{#1}\else $\times 10^{#1}$\fi}
\def\,{\thinspace}
\def\lsim{\mathrel{\raise .4ex\hbox{\rlap{$<$}\lower 1.2ex\hbox{$\sim$}}}}
\def\gsim{\mathrel{\raise .4ex\hbox{\rlap{$>$}\lower 1.2ex\hbox{$\sim$}}}}

\def\simprop{\mathrel{\raise .4ex\hbox{\rlap{$\propto$}\lower 1.2ex\hbox{$\sim$}}}}
\def\deg{\ifmmode^\circ\else$^\circ$\fi}
\def\pdeg{\ifmmode $\setbox0=\hbox{$^{\circ}$}\rlap{\hskip.11\wd0 .}$^{\circ}
          \else \setbox0=\hbox{$^{\circ}$}\rlap{\hskip.11\wd0 .}$^{\circ}$\fi}
\def\arcs{\ifmmode {^{\scriptstyle\prime\prime}}
          \else $^{\scriptstyle\prime\prime}$\fi}
\def\arcm{\ifmmode {^{\scriptstyle\prime}}
          \else $^{\scriptstyle\prime}$\fi}
\newdimen\sa  \newdimen\sb
\def\parcs{\sa=.07em \sb=.03em
     \ifmmode \hbox{\rlap{.}}^{\scriptstyle\prime\kern -\sb\prime}\hbox{\kern -\sa}
     \else \rlap{.}$^{\scriptstyle\prime\kern -\sb\prime}$\kern -\sa\fi}
\def\parcm{\sa=.08em \sb=.03em
     \ifmmode \hbox{\rlap{.}\kern\sa}^{\scriptstyle\prime}\hbox{\kern-\sb}
     \else \rlap{.}\kern\sa$^{\scriptstyle\prime}$\kern-\sb\fi}
\def\ra[#1 #2 #3.#4]{#1\sup{h}#2\sup{m}#3\sup{s}\llap.#4}
\def\dec[#1 #2 #3.#4]{#1\deg#2\arcm#3\arcs\llap.#4}
\def\deco[#1 #2 #3]{#1\deg#2\arcm#3\arcs}
\def\rra[#1 #2]{#1\sup{h}#2\sup{m}}

\def\dots{\relax\ifmmode \ldots\else $\ldots$\fi}
\def\WHzsr{\ifmmode $W\,Hz\mo\,sr\mo$\else W\,Hz\mo\,sr\mo\fi}
\def\mHz{\ifmmode $\,mHz$\else \,mHz\fi}
\def\GHz{\ifmmode $\,GHz$\else \,GHz\fi}
\def\mKs{\ifmmode $\,mK\,s$^{1/2}\else \,mK\,s$^{1/2}$\fi}
\def\muKs{\ifmmode \,\mu$K\,s$^{1/2}\else \,$\mu$K\,s$^{1/2}$\fi}
\def\muKRJs{\ifmmode \,\mu$K$_{\rm RJ}$\,s$^{1/2}\else \,$\mu$K$_{\rm RJ}$\,s$^{1/2}$\fi}
\def\muKHz{\ifmmode \,\mu$K\,Hz$^{-1/2}\else \,$\mu$K\,Hz$^{-1/2}$\fi}
\def\MJysr{\ifmmode \,$MJy\,sr\mo$\else \,MJy\,sr\mo\fi}
\def\MJysrmK{\ifmmode \,$MJy\,sr\mo$\,mK$_{\rm CMB}\mo\else \,MJy\,sr\mo\,mK$_{\rm CMB}\mo$\fi}
\def\microns{\ifmmode \,\mu$m$\else \,$\mu$m\fi}

\def\muK{\ifmmode \,\mu$K$\else \,$\mu$\hbox{K}\fi}
\def\microK{\ifmmode \,\mu$K$\else \,$\mu$\hbox{K}\fi}
\def\muW{\ifmmode \,\mu$W$\else \,$\mu$\hbox{W}\fi}
\def\kms{\ifmmode $\,km\,s$^{-1}\else \,km\,s$^{-1}$\fi}
\def\kmsMpc{\ifmmode $\,\kms\,Mpc\mo$\else \,\kms\,Mpc\mo\fi}
\providecommand{\sorthelp}[1]{}

\def\WMAP{WMAP}

\renewcommand{\L}[0]{\tens{L}}

\def\inv{^{-1}}

\begin{document}

\title{Constraints on the spectral index of polarized synchrotron
  emission from \WMAP\ and Faraday-corrected S-PASS data}
\author{\small
  U.~Fuskeland\inst{1}\thanks{Corresponding author: U.~Fuskeland; \url{unnif@astro.uio.no}}, K.~J.~Andersen\inst{1}, R.~Aurlien\inst{1}, R.~Banerji\inst{1}, M.~Brilenkov\inst{1}, H.~K.~Eriksen\inst{1}, M.~Galloway\inst{1}, E.~Gjerl{\o}w\inst{1}, S.~K.~N{\ae}ss\inst{2}, T.~L.~Svalheim\inst{1}, and I.~K.~Wehus\inst{1}
}
\institute{\small
  Institute of Theoretical Astrophysics, University of Oslo, Blindern, Oslo, Norway\goodbreak
  \and
  Center for Computational Astrophysics, Flatiron Institute, 162 5th Avenue, New York, NY 10010, USA\goodbreak
}

\authorrunning{U. Fuskeland et al.}
\titlerunning{Spectral index of polarized synchrotron emission from S-PASS and \WMAP}
\date{Received 31 January 2020 / Accepted 7 December 2020 in A\&A}

\abstract{We constrain the spectral index of polarized synchrotron
  emission, $\beta_{\mathrm{s}}$, by correlating the recently released 2.3~GHz
  \emph{S-Band Polarization All Sky Survey} (S-PASS) data with the
  23~GHz 9-year \emph{Wilkinson Microwave Anisotropy Probe} (\WMAP)
  sky maps. We subdivide the S-PASS field, which covers the southern
  ecliptic hemisphere, into 95 $15^{\circ}\times15^{\circ}$ regions and
  estimate the spectral index of polarized synchrotron emission within
  each region using a simple but robust $T$--$T$ plot technique. Three
  different versions of the S-PASS data are considered, corresponding
  to: no correction for Faraday rotation; Faraday correction
  based on the rotation measure model presented by the S-PASS team; or
  Faraday correction based on a rotation measure model presented by
  Hutschenreuter and En{\ss}lin. We find that the correlation between
  S-PASS and \WMAP\ is strongest when applying the S-PASS
  model. Adopting this correction model, we find that the mean
  spectral index of polarized synchrotron emission gradually steepens
  from $\beta_{\mathrm{s}}\approx-2.8$ at low Galactic latitudes to
  $\beta_{\mathrm{s}}\approx-3.3$ at high Galactic latitudes, in good agreement
  with previously published results. The flat spectral index at the 
  low Galactic latitudes is likely partly due to depolarization effects. 
  Finally, we consider two special
  cases defined by the BICEP2 and SPIDER fields and obtain mean
  estimates of $\beta_{\textrm{BICEP2}}=-3.22\pm0.06$ and
  $\beta_{\textrm{SPIDER}}=-3.21\pm0.03$, respectively. Adopting the bandpass 
  filtered \WMAP\ 23~GHz sky map to only include angular scales
  between $2^{\circ}$ and $10^{\circ}$ as a spatial template, we
  constrain the root-mean-square synchrotron polarization amplitude to
  be less than $0.03\,\mu\textrm{K}$ ($0.009\,\mu\textrm{K}$) at
  90~GHz (150~GHz) for the BICEP2 field, corresponding roughly to a
  tensor-to-scalar ratio of $r\lesssim0.02$ ($r\lesssim0.005$).
  Very similar constraints are obtained for the SPIDER
  field. A comparison with a similar analysis performed in the $23$-$33$ GHz 
  range suggests a flattening of about 
  $\Delta\beta_s \sim 0.1 \pm 0.2$ from low to higher frequencies, 
  but with no statistical significance due to high uncertainties.

}

\keywords{ISM: general -- Cosmology: observations, polarization,
    cosmic microwave background, diffuse radiation -- Galaxy:
    general}

\maketitle

%\hypersetup{linkcolor=black}
%\tableofcontents
%\hypersetup{linkcolor=red} 

\section{Introduction}
\label{sec:introduction}

The field of observational cosmology has undergone a dramatic
transformation in recent decades. The main driving force behind
these developments has been rapidly improving instrumentation across
the electromagnetic spectrum. This holds particularly true for
measurements in the microwave range, which are essential for mapping
the cosmic microwave background (CMB), an afterglow from the Big
Bang. Such observations constrain cosmological parameters and models
to sub-percent accuracy, the most prominent demonstration of which 
has been the European Space Agency’s (ESA) \Planck\ satellite mission
\citep{planck2016-l01, planck2016-l06}.

While detailed measurements of the CMB temperature and polarization
fluctuations have already transformed cosmology, such measurements 
further hold the promise of providing a unique window into the physics
during the first tiny fraction of a second following the Big
Bang. Specifically, according to the current standard cosmological
concordance model, a quantum mechanical process called inflation
\citep[see, e.g.,][and references therein]{liddle:1999} took place
shortly after the Big Bang, during which the effective length scale of
the universe increased by a factor of $10^{28}$ or more during some
$10^{-34}$ seconds. As a result of this process, space was
violently stretched, and a background of so-called
primordial gravitational waves was excited. These gravitational waves
later warped spacetime during the epoch of recombination, stretching
space in one direction and compressing it in the orthogonal direction,
and created a particular unique signal in the CMB field that today can
be observed in the form of so-called B-mode polarization \citep[e.g.,][]{zaldarriaga:1997}.

Robustly detecting the polarization signature of these primordial
B-modes would provide cosmologists with a unique opportunity to constrain 
physics at the Planck scale. Unfortunately, the expected amplitude of the 
signal is very small for currently viable theories, ranging up to no more 
than 100\,nK on large angular scales, and probably significantly
less \citep{bicep2}. A wide range of these models are within reach, 
and even if these amplitudes are within the capabilities 
of modern detectors in terms of raw noise performance, another issue
complicates the picture considerably, namely foreground emission from
interstellar particles situated within the Milky Way. In particular,
relativistic electrons moving within the Galactic magnetic field emit
polarized synchrotron emission, whereas small vibrating dust grains
aligned by the same magnetic field emit polarized thermal
emission. Both of these foreground signals are very likely orders of
magnitude brighter than the primordial gravitational wave signal on
large angular scales \citep{planck2016-l04}.

Robustly distinguishing between the primordial and the local
polarization signals is among the key challenges of modern CMB
cosmology, and great efforts are being made to both establish
observational constraints on the various effects and develop
efficient computational and statistical methods to analyze the
resulting data \citep[e.g.,][and references therein]{leach:2008}. So
far, stronger constraints have been derived for polarized
thermal dust than for synchrotron, largely because the detectors
needed to probe the relevant frequency range are smaller, cheaper, and
more sensitive than the corresponding detectors required to probe
synchrotron. Among the best examples of this are the \Planck\ 217-
and 353-GHz channels, which have revolutionized our understanding of
polarized thermal dust in the CMB frequency range \citep{planck_PIP_XIX}.

Until very recently, the Wilkinson Microwave Anisotropy Probe (\WMAP;
\citealp{bennett2012}) 23~GHz and \Planck\ 30~GHz \citep{planck2016-l02}
frequency channels provided the strongest constraints on polarized
synchrotron emission. Subsequently, in March 2019 the first sky maps from
the \emph{S-Band Polarization All Sky Survey} (S-PASS;
\citealp{Carretti:2019}) were publicly released, observed at 2.3~GHz. 
Due to the spectral energy density power law relation of synchrotron 
emission, the signal at 2.3~GHz is in fact about 1000 times stronger 
than at 23~GHz, thus S-PASS provides a clear image of synchrotron 
emission in both intensity and polarization.
The S-PASS map covers most of the southern celestial
sphere (dec < $-1^{\circ}$), for a total sky fraction of 48.7\%. A
total of 98.6\% of all pixels have a reported polarization
signal-to-noise higher than three. This makes S-PASS an excellent
complement to \WMAP\ and \Planck, and jointly they should provide strong
constraints on polarized synchrotron emission in the microwave
regime. Indeed, an early analysis of this type has already been
presented by \citet{krachmalnicoff:2019}.

However, while the S-PASS data contain a wealth of information on
synchrotron emission, their utility is significantly complicated by
Faraday rotation \citep[e.g.,][]{beck:2013}. First discovered by
Michael Faraday in 1845, this effect causes the rotation of the plane
of polarization of an electromagnetic wave in the presence of a
magnetic field. The effect is proportional to the strength
of the magnetic field and the integrated electron density, 
as well as to the square of the wavelength of the wave. 
In an astrophysical setting, the Faraday
rotation effect is therefore stronger for low frequencies and at low
Galactic latitudes. For instance, while the magnitude of the effect is
typically a few degrees at 23~GHz, it can be many hundreds of degrees at
2.3~GHz along the Galactic plane. Even at high Galactic latitudes, it
can be several tens of degrees at this low frequency.

The magnitude of the Faraday rotation effect is typically quantified
in terms of the rotation measure ($\textrm{RM}$), which is simply the
proportionality constant that scales the square of the wavelength. Several
models\footnote{In the following, a "rotation measure model" refers to a 
numerical approximation to the true rotation measure that may or may not 
be constrained by observations. } 
have been derived for the rotation measure, and in this paper
we will consider and quantitatively compare three different
models. The first is simply assuming no Faraday rotation at all, namely
$\textrm{RM}=0$. This serves as a baseline that allows us to assess the impact 
of the Faraday rotation effect. Our second model is that derived by the S-PASS
team as part of the data release \citep{Carretti:2019}. This model was
derived as a joint fit to the S-PASS, \WMAP\ 23~GHz, and \Planck\ 30~GHz
data sets. Our third and final model is that derived by
\citet{Hutschenreuter:2019} through a Bayesian analysis of
extra-galactic point sources and the \Planck\ free-free map. 

Throughout this paper we use a convention for polarization angles
(PAs) where PA is $0^{\circ}$ for vectors pointing north and increases
westward. The same convention is used in experiments such as \WMAP\ and
\Planck and differs from the International Astronomical Union (IAU) 
convention used in the S-PASS experiment where the PA increases eastward.  

The rest of the paper is organized as follows. In Sect.~\ref{sec:data}
we briefly describe the data used in this paper, and in
Sect.~\ref{sec:faraday_sky} we provide details on the Faraday rotation
models we employ. The algorithms used to estimate the spectral index
are described in Sect.~\ref{sec:methods}, while the main results are
presented in Sect.~\ref{sec:results}. In Sect.~\ref{sec:special_cases} we
consider two special cases, namely the BICEP2 and SPIDER fields, both
of which are covered by S-PASS. Finally, we
conclude in Sect.~\ref{sec:conclusions}. 

\begin{figure*}

  \centering
\subfigure{
\includegraphics[width=0.49\linewidth]{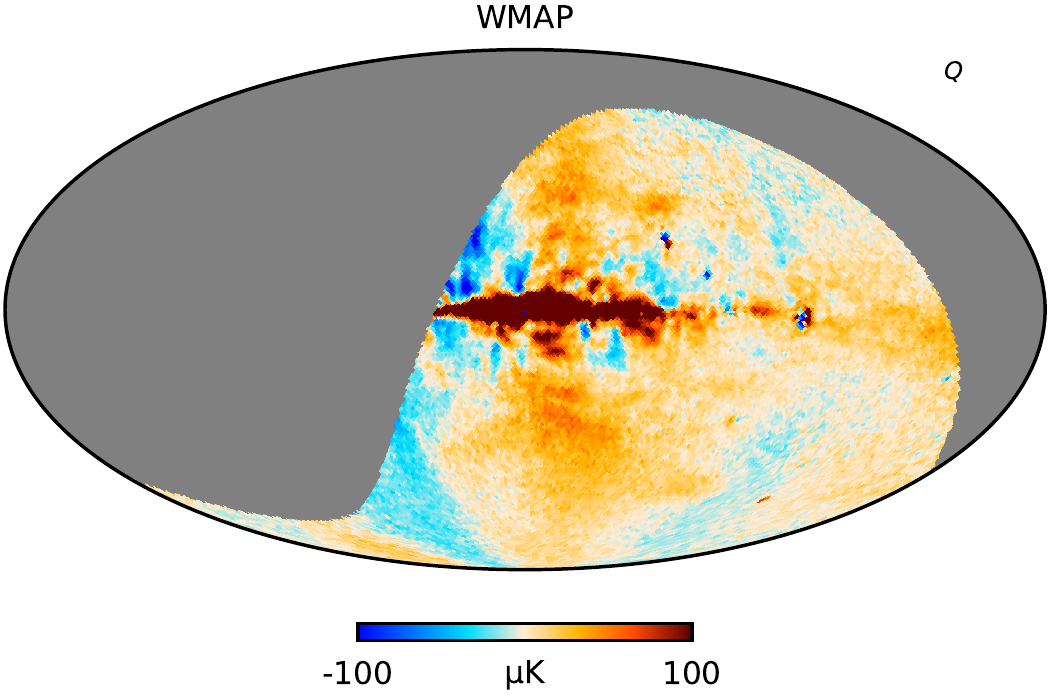}
\includegraphics[width=0.49\linewidth]{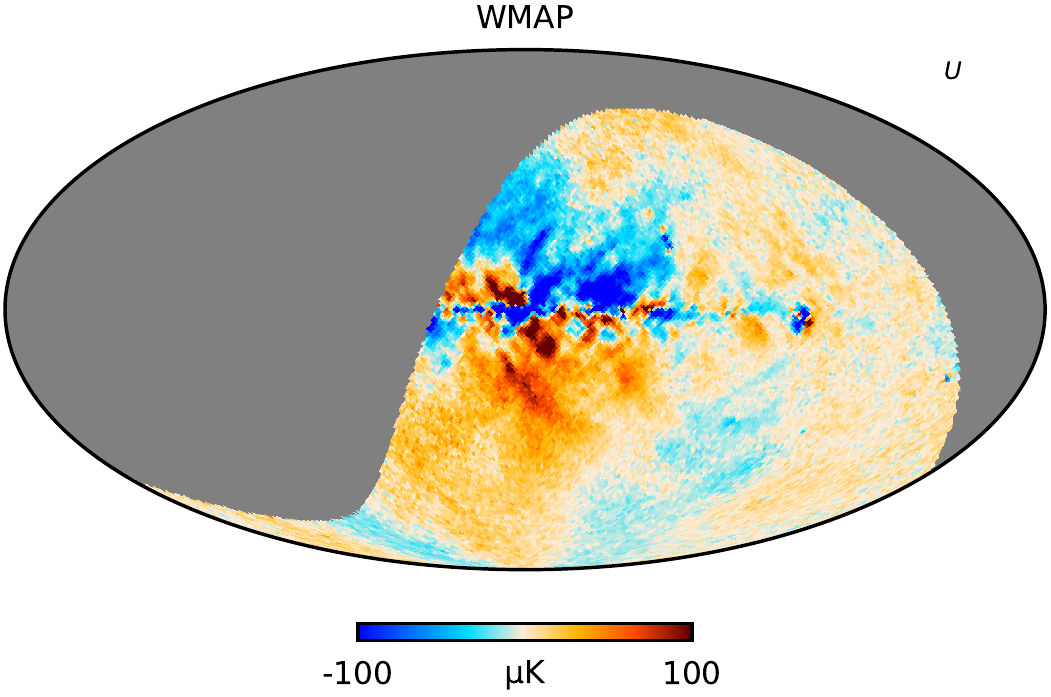}}

%\vspace{-0.33in}                                                                                                     

\subfigure{
\includegraphics[width=0.49\linewidth]{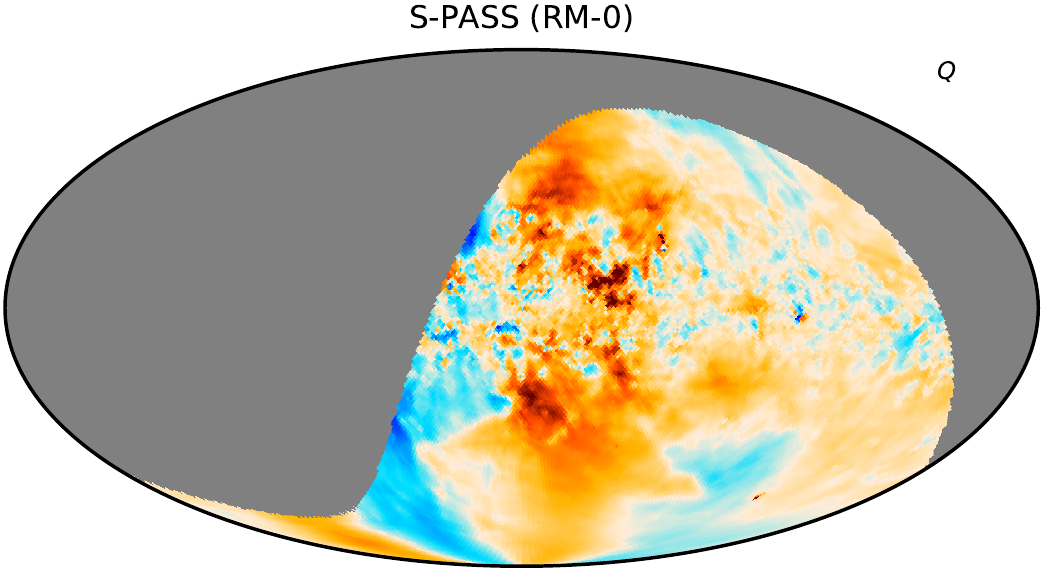}
\includegraphics[width=0.49\linewidth]{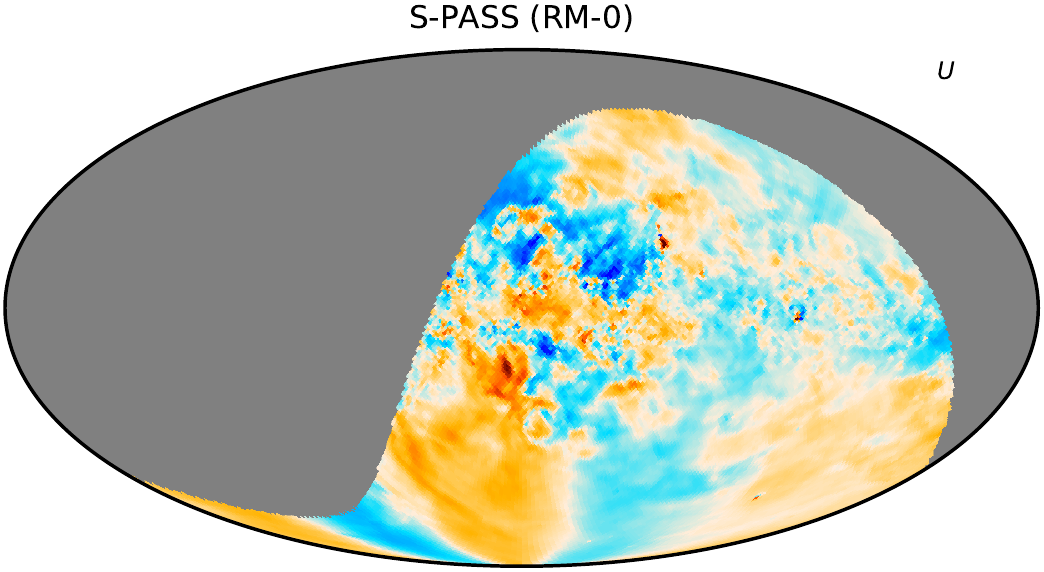}}

\subfigure{
\includegraphics[width=0.49\linewidth]{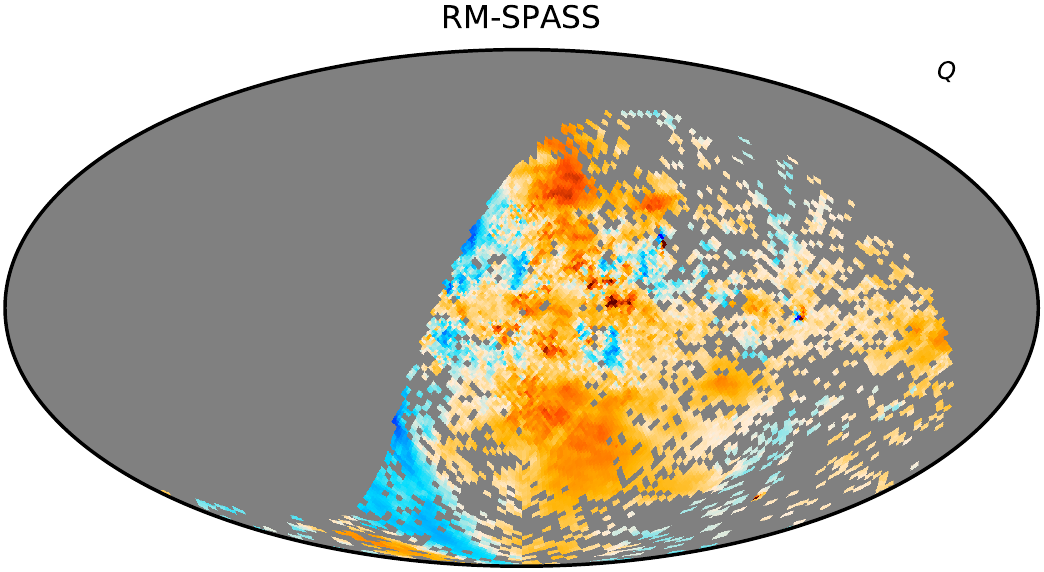}
\includegraphics[width=0.49\linewidth]{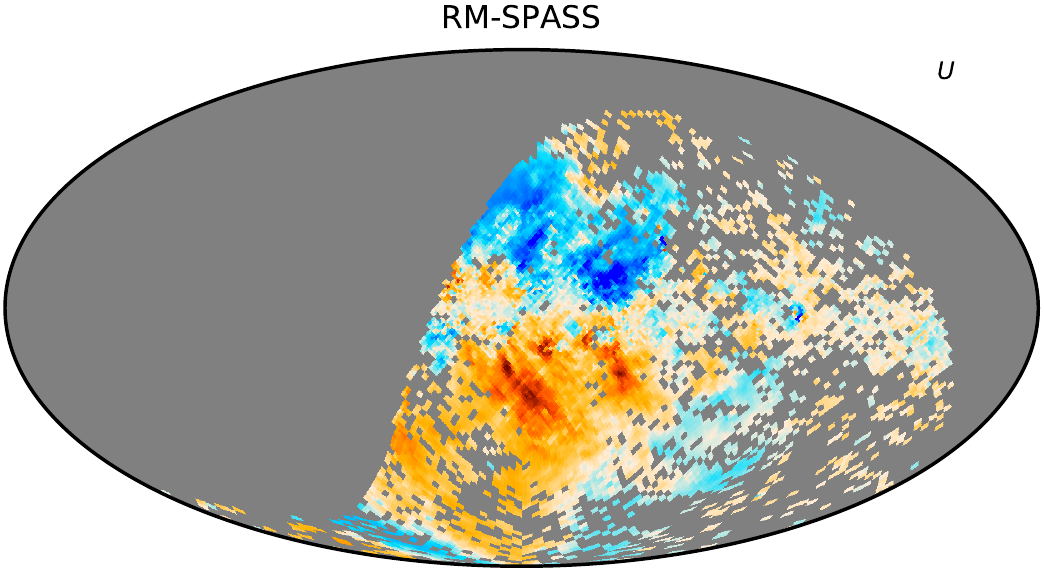}}

\subfigure{
\includegraphics[width=0.49\linewidth]{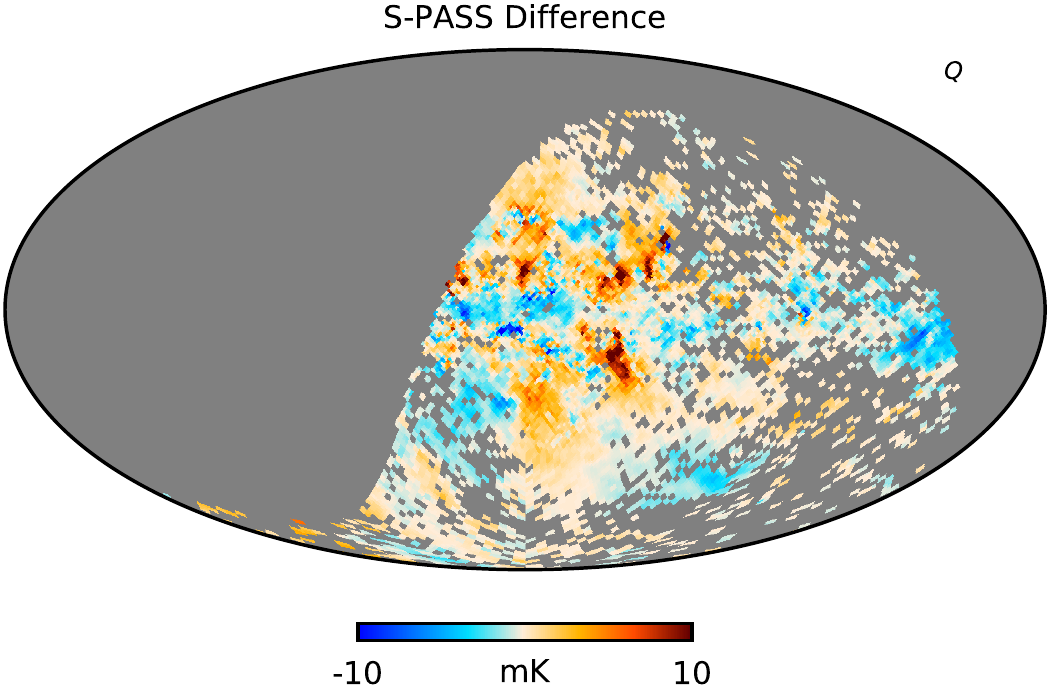}
\includegraphics[width=0.49\linewidth]{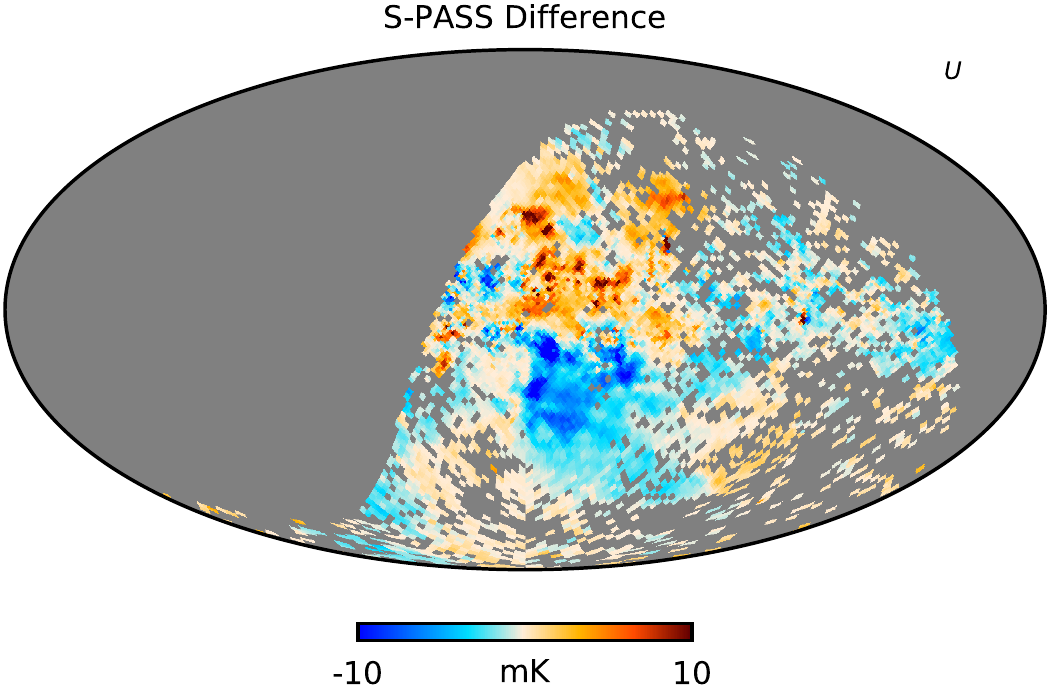}}

\caption{Comparison of the main sky maps used in this analysis. From top
  to bottom, rows show the 23~GHz \WMAP\ map (\emph{top row}); the raw
  2.3~GHz S-PASS (\emph{second row}); the same S-PASS sky map, but
  corrected for Faraday rotation using the RM-SPASS \citep{Carretti:2019}
  model (\emph{third row}); and the difference between the raw and the
  corrected S-PASS maps (\emph{bottom row}). Left and right columns
  show the Stokes $Q$ and $U$ components, respectively. All maps are
  smoothed to a common angular resolution of $1^{\circ}$ FWHM, and all
  maps are plotted in brightness (Rayleigh-Jeans) temperature units.}
\label{fig:spass}
\end{figure*}

\section{S-PASS and \WMAP\ data}
\label{sec:data}

The main goal of this paper is to estimate the synchrotron spectral
index exploiting the statistical power of
the recently released S-PASS sky map. To complement this map, we
choose the \WMAP\ 23~GHz sky map \citep{bennett2012}, simply because it
has higher signal-to-noise to polarized synchrotron emission compared
to other available alternatives, most notably the \Planck\ 30~GHz
channel \citep{planck2016-l01}.

First, we note that the S-PASS data were collected with the Parkes
radio telescope, and therefore covers the southern celestial sky at
dec $<-1^\circ$, whereas the \WMAP\ data are all-sky. As such, we
apply an analysis mask, and consider only pixels within the S-PASS
coverage, for a total of 48.7\,\% of the sky.

Second, the S-PASS sky map has a native angular resolution of 8.9'
full width at half maximum (FWHM),
whereas the \WMAP\ 23~GHz sky map has a resolution of 53'
FWHM. Further, the two maps are pixelized on different grids, as
S-PASS is defined on a HEALPix\footnote{http://healpix.jpl.nasa.gov}
\citep{gorski2005} grid with $N_{\textrm{side}}=1024$ (3.4' pixel
size), while \WMAP\ is defined on an $N_{\textrm{side}}=512$ (6.7' pixel
size) grid. Our analysis requires both maps to be smoothed to a common
angular resolution and pixelized with the same grid, and we therefore
adopt a common resolution of $1^{\circ}$ FWHM and $N_{\mathrm{side}}=64$ (55'
pixel size). Such a coarse pixel size ensures that
neighboring pixels are only weakly correlated, and since no subsequent
spherical harmonics transforms are involved in the analysis, operating
with non-bandwidth limited maps, (i.e., corresponding to non-Nyquist 
sampling limited maps in the flat space case), is not a concern for 
this particular analysis.

For S-PASS we adopted an effective frequency of 2.303~GHz
\citep{Carretti:2019}, while for the \WMAP\ 23~GHz sky maps we adopted
an effective frequency of 22.45~GHz, corresponding to the effective
frequency of a synchrotron spectrum scaling as $\nu^{-3}$ integrated
over the \WMAP\ bandpass \citep{page2003a}. At the low frequencies
discussed in this paper, other sources of polarized
emission (thermal dust being the dominant one) have a signal of about
one percent of that of the synchrotron emission at a frequency of
23~GHz, while at the S-PASS frequency they are totally negligible. We
therefore assume that both maps contain only polarized synchrotron
emission and noise.

The top row of Fig.~\ref{fig:spass} shows the \WMAP\ sky map in the
S-PASS field, smoothed to $1^{\circ}$ FWHM, while the second row 
shows the corresponding S-PASS sky map. 
Left and right columns show the Stokes $Q$ and $U$
parameters. As already noted in the introduction, we adopt the same
convention for the polarization angles as \WMAP\ and \Planck. This is
different from the S-PASS convention, for which
the polarization angle increases eastward. To account for this
difference, we multiply the S-PASS Stokes $U$ parameter by $-1$.

By eye, one can clearly see a strong correlation between the S-PASS
and \WMAP\ sky maps at high Galactic latitudes. However, at low
latitudes there are major differences. Most notably, while
\WMAP\ exhibits a strong $Q$ component, indicating a structured
magnetic field oriented parallel to the Galactic plane, the S-PASS map
has virtually no signal in the Galactic plane. This is a typical
signature of Faraday rotation, which effectively rotates the
polarization angle for a given emission source through a random angle
before arriving at our location in the Milky Way. When integrating
over many such sources, each with a random angle depending on its
distance, the net sum is dramatically decreased. This is often
referred to as ``Faraday depolarization,'' and is likely to flatten the spectral index in proximity to the Galactic plane.

\begin{figure}
  \begin{center}
    \includegraphics[width=\linewidth]{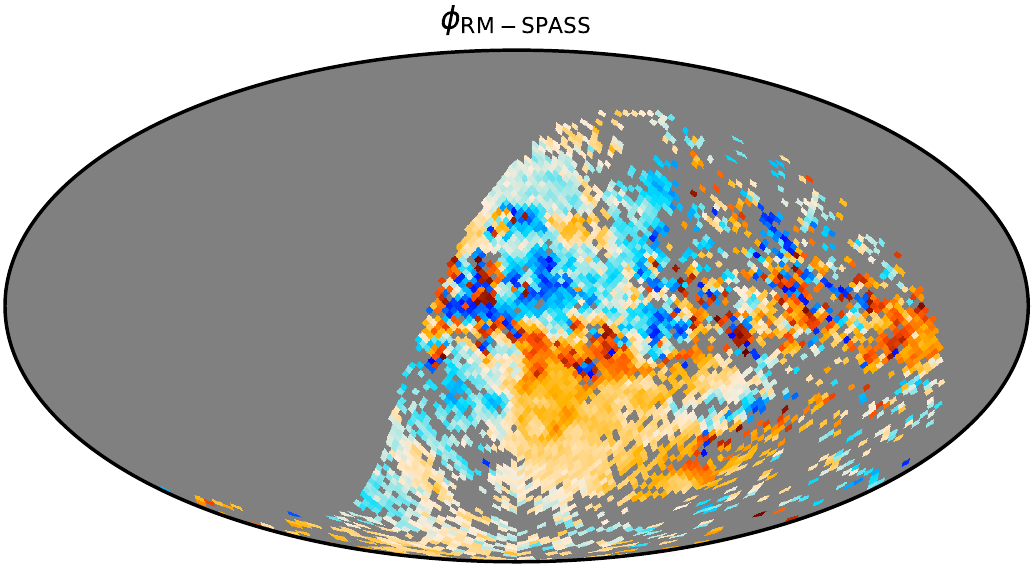}
    \includegraphics[width=\linewidth]{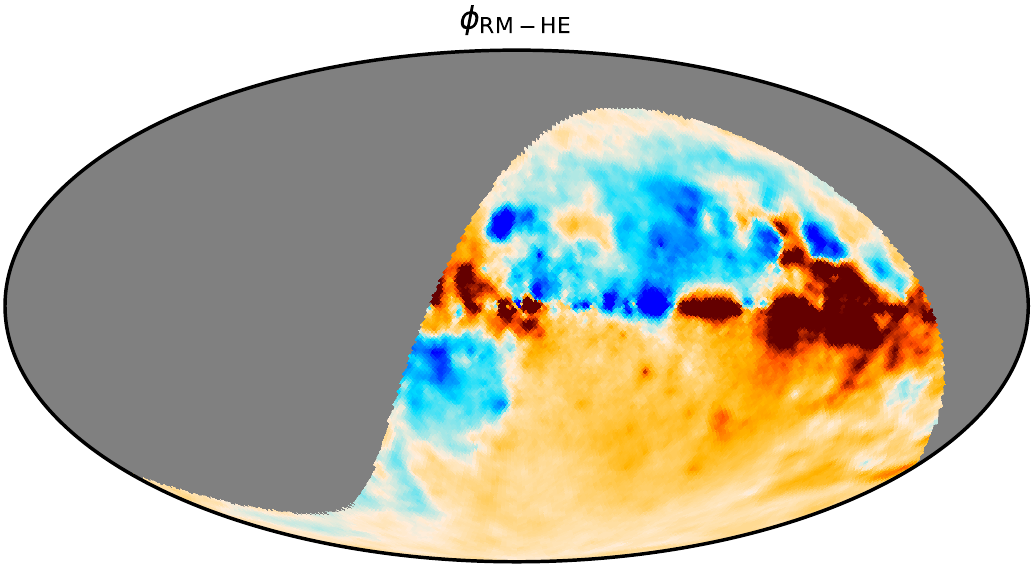}
    \includegraphics[width=\linewidth]{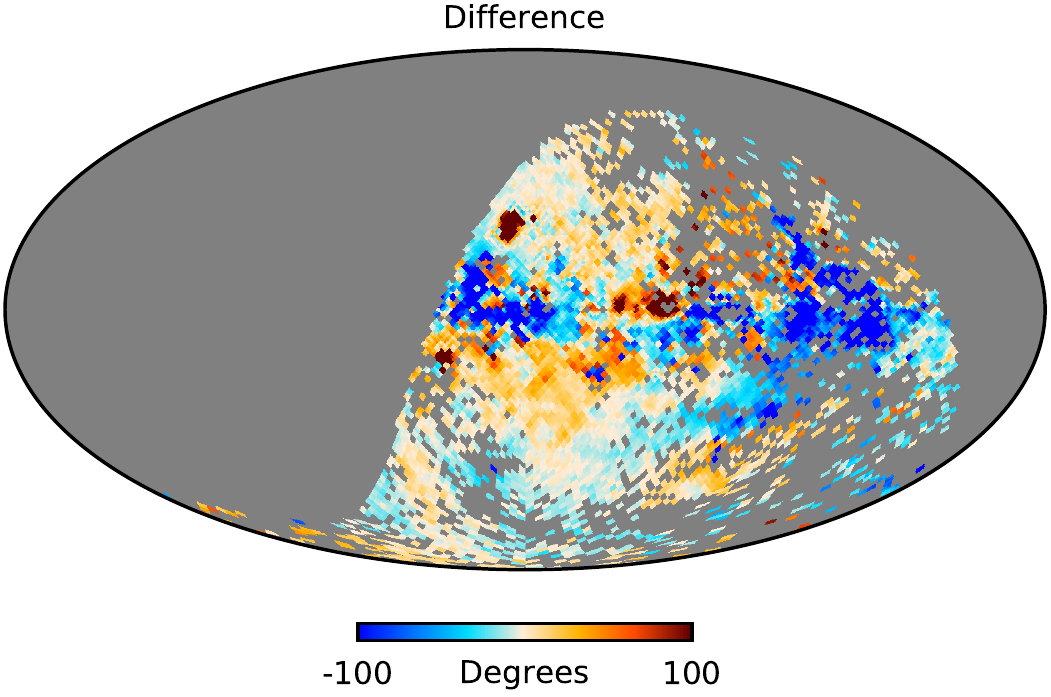}
  \end{center}
  \caption{Predicted Faraday rotation angle, $\phi_{\mathrm{RM}}$, at
    2.3~GHz for the two models considered in this paper. The top panel
    shows the RM-SPASS model \citep{Carretti:2019}, while the middle
    panel shows the RM-HE model \citep{Hutschenreuter:2019}. The bottom
    panel shows the difference between the two models. All maps are
    shown in units of degrees. }
  \label{fig:faraday_degspass}
\end{figure}

%%%%%%%%%%%%%%%%%%%%
\section{Faraday rotation models and corrections}
\label{sec:faraday_sky}

The total polarization angle $\phi_\lambda$ of linearly polarized light
due to Faraday rotation, $\phi_{\mathrm{RM}}$, can be written as \citep[e.g.,][]{beck:2013}
\begin{equation}
\phi_\lambda = \phi_0 + \phi_{\mathrm{RM}} = \phi_0 + \mathrm{RM}\, \lambda^2,
\label{eq:rm}
\end{equation}
where $\phi_0$ is the intrinsic polarization angle of the source,
$\mathrm{RM}$ is the rotation measure in units of rad\,m$^{-2}$, and
$\lambda$ is the wavelength of the radiation. We consider two
nontrivial models for the rotation measure in this paper constrained by 
direct measurements. 
The first model was presented as part of the S-PASS data release, and was
derived through a joint fit to the S-PASS, \WMAP, and \Planck\ data
\citep{Carretti:2019}. The second model was presented by
\citet{Hutschenreuter:2019}, which used a combination of
extra-galactic point sources and the \Planck\ Commander free-free map
to constrain the Galactic rotation measure within a Bayesian framework. 
Since this is measured using extra-galactic sources, it might not be an 
appropriate choice to correct the polarization angles of the diffuse 
emission considered in this paper. Nevertheless, we have chosen to 
include this template in the analysis. 
For completeness, we also consider the trivial case in which no correction for
Faraday rotation is applied, namely $\mathrm{RM}=0$. We will refer to
these three models as RM-SPASS, RM-HE, and RM-0, respectively.

Figure~\ref{fig:faraday_degspass} compares RM-SPASS (top panel) and
RM-HE (middle panel) in terms of the predicted rotation angle at
2.3~GHz. The bottom panel shows the difference between the two
models. For both models we see that the predicted rotation angle is
quite large for low Galactic latitudes, and small relative errors can
therefore give large biases in a map that is rotated using these
templates. It is also worth noting that the difference between the two
models is substantial not only at low Galactic latitudes, but also at
intermediate and high latitudes, at the level of tens of degrees.

We note that the RM-SPASS model has many missing pixels within the S-PASS
region. These are pixels for which the S-PASS collaboration
considered the error on the RM or the difference in angle between WMAP 
and PLANCK angle maps too large to be reliable, and therefore did not 
provide an estimate. We also exclude these pixels in all subsequent 
analyses involving this model.

Considering that the center frequencies of the two data sets in
question in this paper are 2.3 and 23~GHz, the
predicted Faraday corrections for \WMAP\ are roughly 100 times smaller
than for S-PASS. As such, they only reach a few degrees in the central
Galactic plane and are negligible at high latitudes. For this reason,
we applied no corrections to \WMAP.

\begin{figure}[t]
  \begin{center}
    \includegraphics[width=\linewidth]{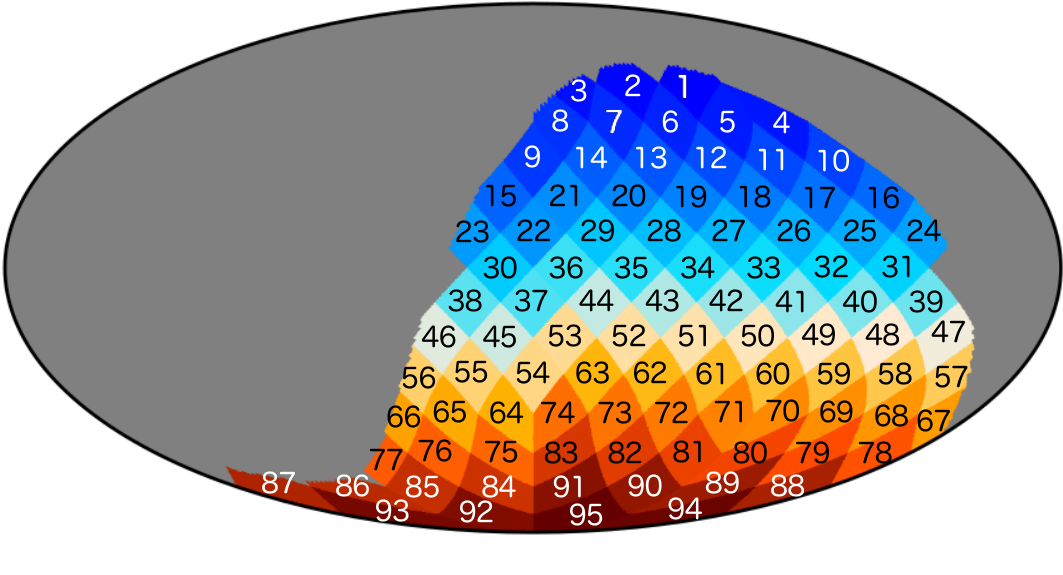}
    \includegraphics[width=\linewidth]{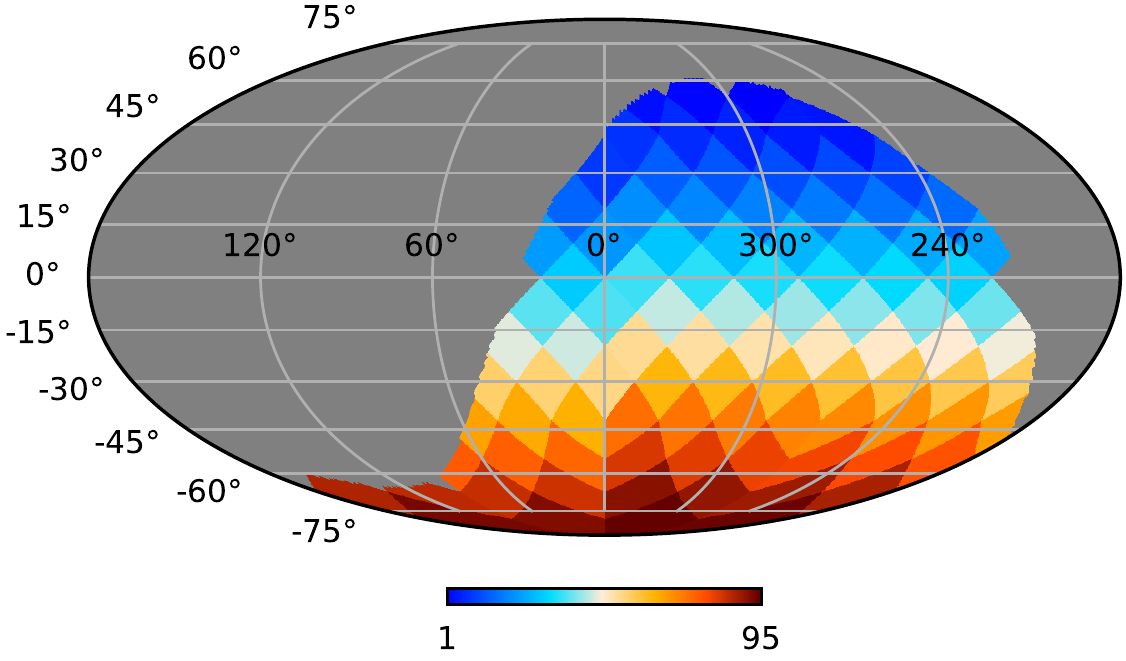}
  \end{center}
  \caption{Subdivision of the S-PASS region according to an
    $N_{\textrm{side}}=4$ HEALPix grid. A total of 95 regions 
    have more than 50\,\% coverage within the S-PASS mask, and 
    these form the primary basis for spatial localization in this paper.
    The top panel shows the region numbers, while the bottom panel shows 
    a grid with galactic coordinates. The color range denote region number.
  }
  \label{fig:regions}
\end{figure}

Based on these models, we produced Faraday-corrected versions of the
S-PASS sky map by performing the following rotation pixel-by-pixel,
\begin{equation}
  \left[
    \begin{array}{c}
      \hat{Q} \\
      \hat{U}
    \end{array}
    \right]
  =
  \left[
    \begin{array}{cc}
     \phantom{-}\cos 2\phi_{\mathrm{RM}} &  \sin 2\phi_{\mathrm{RM}} \\
      -\sin 2\phi_{\mathrm{RM}} &  \cos 2\phi_{\mathrm{RM}}
    \end{array}
    \right]  
  \left[
    \begin{array}{c}
      Q \\
      U
    \end{array}
    \right].
\end{equation}
Here $(Q,U)$ denote the observed Stokes parameters, and $(\hat{Q},\hat{U})$
represent the Faraday-corrected Stokes parameters. The sign
is chosen such that the correction corresponds to a negative rotation
angle compared to those predicted by the model. This takes into account 
that the two RM models follow the IAU convention of the polarization angles.

If the adopted model represents an accurate estimate of the true sky,
these Faraday corrections should improve the correlations between the
S-PASS and \WMAP\ data. We therefore compute the mean Pearson
correlation coefficient between these two data sets for each of the
three models as follows.

\begin{table}[t]  % Table 1
\begingroup
\newdimen\tblskip \tblskip=5pt
\caption{Synchrotron spectral index for each region \label{tab:betas}}
\vskip -6mm
\scriptsize
%\footnotesize
\setbox\tablebox=\vbox{
\newdimen\digitwidth
\setbox0=\hbox{\rm 0}
\digitwidth=\wd0
\catcode`*=\active
\def*{\kern\digitwidth}
\newdimen\signwidth
\setbox0=\hbox{+}
\signwidth=\wd0
\catcode`!=\active
\def!{\kern\signwidth}
\newdimen\decimalwidth
\setbox0=\hbox{.}
\decimalwidth=\wd0
\catcode`@=\active
\def@{\kern\signwidth}
\halign{\hbox to 0.9in{#\leaderfil}\tabskip=1.0em&
    \hfil#\hfil\tabskip=0.4em&
    \hfil#\hfil\tabskip=0.2em&
    \hfil#\hfil\tabskip=0.2em&
    \hfil#\hfil\tabskip=0em\cr
\noalign{\doubleline}
\omit\hfil Region\hfil &Latitude&Longitude& Un-corr.&Faraday-corr.\cr 
\noalign{\vskip 5pt\hrule\vskip 5pt}
\noalign{\vskip 3pt}
%\multispan5{\bf \boldmath S-PASS}\hfil\cr 
\noalign{\vskip 3pt}
\hglue 0em 1& $ 53^{\circ} $& $ 287^{\circ} $& $ -3.17\pm0.10$& $ -$\cr
\hglue 0em 2& $ 53^{\circ} $& $ 315^{\circ} $& $ -3.21\pm0.11$& $ -$\cr
\hglue 0em 3& $ 50^{\circ} $& $ 340^{\circ} $& $ -3.50\pm0.15$& $ -3.27\pm0.10$\cr
\hglue 0em 5& $ 42^{\circ} $& $ 280^{\circ} $& $ -3.22\pm0.10$& $ -$\cr
\hglue 0em 6& $ 42^{\circ} $& $ 303^{\circ} $& $ -$& $ -3.29\pm0.09$\cr
\hglue 0em 7& $ 42^{\circ} $& $ 327^{\circ} $& $ -3.26\pm0.08$& $ -3.27\pm0.09$\cr
\hglue 0em 8& $ 41^{\circ} $& $ 349^{\circ} $& $ -3.27\pm0.18$& $ -3.25\pm0.06$\cr
\hglue 0em 11& $ 30^{\circ} $& $ 270^{\circ} $& $ -$& $ -3.12\pm0.04$\cr
\hglue 0em 12& $ 30^{\circ} $& $ 293^{\circ} $& $ -$& $ -3.25\pm0.13$\cr
\hglue 0em 13& $ 30^{\circ} $& $ 315^{\circ} $& $ -3.29\pm0.21$& $ -3.19\pm0.03$\cr
\hglue 0em 14& $ 30^{\circ} $& $ 338^{\circ} $& $ -3.46\pm0.15$& $ -3.25\pm0.03$\cr
\hglue 0em 15& $ 20^{\circ} $& $ 11^{\circ} $& $ -$& $ -3.12\pm0.13$\cr
\hglue 0em 17& $ 20^{\circ} $& $ 259^{\circ} $& $ -$& $ -3.09\pm0.06$\cr
\hglue 0em 18& $ 20^{\circ} $& $ 281^{\circ} $& $ -3.27\pm0.26$& $ -3.15\pm0.08$\cr
\hglue 0em 19& $ 20^{\circ} $& $ 304^{\circ} $& $ -$& $ -3.07\pm0.06$\cr
\hglue 0em 20& $ 20^{\circ} $& $ 326^{\circ} $& $ -$& $ -3.22\pm0.07$\cr
\hglue 0em 21& $ 20^{\circ} $& $ 349^{\circ} $& $ -$& $ -3.19\pm0.14$\cr
\hglue 0em 22& $ 10^{\circ} $& $ 360^{\circ} $& $ -$& $ -2.80\pm0.14$\cr
\hglue 0em 23& $ 9^{\circ} $& $ 21^{\circ} $& $ -$& $ -3.02\pm0.04$\cr
\hglue 0em 26& $ 10^{\circ} $& $ 270^{\circ} $& $ -$& $ -3.07\pm0.15$\cr
\hglue 0em 27& $ 10^{\circ} $& $ 293^{\circ} $& $ -3.36\pm0.11$& $ -3.13\pm0.06$\cr
\hglue 0em 28& $ 10^{\circ} $& $ 315^{\circ} $& $ -$& $ -3.16\pm0.03$\cr
\hglue 0em 29& $ 10^{\circ} $& $ 338^{\circ} $& $ -$& $ -3.26\pm0.13$\cr
\hglue 0em 32& $ 0^{\circ} $& $ 259^{\circ} $& $ -$& $ -2.74\pm0.07$\cr
\hglue 0em 33& $ 0^{\circ} $& $ 281^{\circ} $& $ -$& $ -2.97\pm0.07$\cr
\hglue 0em 34& $ 0^{\circ} $& $ 304^{\circ} $& $ -$& $ -2.77\pm0.14$\cr
\hglue 0em 35& $ 0^{\circ} $& $ 326^{\circ} $& $ -$& $ -2.82\pm0.13$\cr
\hglue 0em 36& $ 0^{\circ} $& $ 349^{\circ} $& $ -$& $ -2.36\pm0.20$\cr
\hglue 0em 37& $ -10^{\circ} $& $ 360^{\circ} $& $ -$& $ -2.93\pm0.08$\cr
\hglue 0em 38& $ -10^{\circ} $& $ 23^{\circ} $& $ -$& $ -3.08\pm0.10$\cr
\hglue 0em 39& $ -10^{\circ} $& $ 225^{\circ} $& $ -$& $ -3.59\pm0.12$\cr
\hglue 0em 41& $ -10^{\circ} $& $ 270^{\circ} $& $ -$& $ -2.96\pm0.21$\cr
\hglue 0em 43& $ -10^{\circ} $& $ 315^{\circ} $& $ -$& $ -2.89\pm0.34$\cr
\hglue 0em 44& $ -10^{\circ} $& $ 338^{\circ} $& $ -$& $ -3.04\pm0.11$\cr
\hglue 0em 45& $ -20^{\circ} $& $ 11^{\circ} $& $ -$& $ -3.09\pm0.10$\cr
\hglue 0em 46& $ -20^{\circ} $& $ 33^{\circ} $& $ -$& $ -3.18\pm0.05$\cr
\hglue 0em 48& $ -20^{\circ} $& $ 236^{\circ} $& $ -$& $ -3.27\pm0.09$\cr
\hglue 0em 50& $ -20^{\circ} $& $ 281^{\circ} $& $ -$& $ -3.49\pm0.05$\cr
\hglue 0em 52& $ -20^{\circ} $& $ 326^{\circ} $& $ -$& $ -3.29\pm0.04$\cr
\hglue 0em 53& $ -20^{\circ} $& $ 349^{\circ} $& $ -$& $ -3.17\pm0.06$\cr
\hglue 0em 54& $ -30^{\circ} $& $ 360^{\circ} $& $ -$& $ -3.17\pm0.14$\cr
\hglue 0em 55& $ -30^{\circ} $& $ 23^{\circ} $& $ -3.35\pm0.19$& $ -3.13\pm0.10$\cr
\hglue 0em 56& $ -32^{\circ} $& $ 43^{\circ} $& $ -3.41\pm0.25$& $ -3.35\pm0.08$\cr
\hglue 0em 58& $ -30^{\circ} $& $ 225^{\circ} $& $ -3.29\pm0.12$& $ -$\cr
\hglue 0em 59& $ -30^{\circ} $& $ 248^{\circ} $& $ -$& $ -3.28\pm0.11$\cr
\hglue 0em 62& $ -30^{\circ} $& $ 315^{\circ} $& $ -3.47\pm0.17$& $ -3.30\pm0.08$\cr
\hglue 0em 63& $ -30^{\circ} $& $ 338^{\circ} $& $ -$& $ -3.10\pm0.15$\cr
\hglue 0em 64& $ -42^{\circ} $& $ 10^{\circ} $& $ -3.25\pm0.21$& $ -3.24\pm0.11$\cr
\hglue 0em 65& $ -42^{\circ} $& $ 33^{\circ} $& $ -3.35\pm0.15$& $ -3.14\pm0.08$\cr
\hglue 0em 66& $ -45^{\circ} $& $ 54^{\circ} $& $ -$& $ -3.35\pm0.22$\cr
\hglue 0em 70& $ -42^{\circ} $& $ 260^{\circ} $& $ -$& $ -3.17\pm0.02$\cr
\hglue 0em 71& $ -42^{\circ} $& $ 280^{\circ} $& $ -$& $ -3.21\pm0.12$\cr
\hglue 0em 73& $ -42^{\circ} $& $ 327^{\circ} $& $ -3.37\pm0.28$& $ -3.23\pm0.21$\cr
\hglue 0em 74& $ -42^{\circ} $& $ 350^{\circ} $& $ -$& $ -3.21\pm0.10$\cr
\hglue 0em 76& $ -54^{\circ} $& $ 45^{\circ} $& $ -3.38\pm0.14$& $ -3.24\pm0.09$\cr
\hglue 0em 77& $ -58^{\circ} $& $ 74^{\circ} $& $ -3.36\pm0.11$& $ -3.31\pm0.22$\cr
\hglue 0em 80& $ -55^{\circ} $& $ 255^{\circ} $& $ -$& $ -3.19\pm0.05$\cr
\hglue 0em 81& $ -55^{\circ} $& $ 285^{\circ} $& $ -$& $ -3.24\pm0.05$\cr
\hglue 0em 82& $ -54^{\circ} $& $ 315^{\circ} $& $ -$& $ -3.30\pm0.05$\cr
\hglue 0em 83& $ -55^{\circ} $& $ 345^{\circ} $& $ -3.28\pm0.28$& $ -3.22\pm0.07$\cr
\hglue 0em 84& $ -67^{\circ} $& $ 22^{\circ} $& $ -3.23\pm0.10$& $ -3.30\pm0.19$\cr
\hglue 0em 85& $ -67^{\circ} $& $ 68^{\circ} $& $ -3.30\pm0.03$& $ -3.34\pm0.14$\cr
\hglue 0em 86& $ -69^{\circ} $& $ 109^{\circ} $& $ -3.42\pm0.09$& $ -$\cr
\hglue 0em 89& $ -67^{\circ} $& $ 248^{\circ} $& $ -3.13\pm0.15$& $ -3.12\pm0.04$\cr
\hglue 0em 90& $ -67^{\circ} $& $ 292^{\circ} $& $ -$& $ -3.29\pm0.12$\cr
\hglue 0em 91& $ -67^{\circ} $& $ 338^{\circ} $& $ -3.32\pm0.10$& $ -3.31\pm0.07$\cr
\hglue 0em 92& $ -79^{\circ} $& $ 45^{\circ} $& $ -3.36\pm0.03$& $ -3.36\pm0.02$\cr
\hglue 0em 93& $ -79^{\circ} $& $ 135^{\circ} $& $ -3.40\pm0.14$& $ -3.40\pm0.02$\cr
\hglue 0em 94& $ -79^{\circ} $& $ 225^{\circ} $& $ -3.36\pm0.14$& $ -3.33\pm0.06$\cr
\hglue 0em 95& $ -79^{\circ} $& $ 315^{\circ} $& $ -3.38\pm0.07$& $ -3.38\pm0.05$\cr
\noalign{\vskip 3pt}
\hglue 0em Mean S-PASS& $-$& $-$ &$-3.32\pm0.02$&$ -3.24\pm0.01$\cr
\hglue 0em Standard deviation S-PASS& $-$& $-$ &$ \phantom{-88.88\pm}0.09$&$ \phantom{-88.88\pm}0.19$\cr
%\multispan5{\bf \boldmath $EE$}\hfil\cr 
\noalign{\vskip 3pt}
\hglue 0em BICEP2& $ -57^{\circ}$& $ 315^{\circ}$& $ -3.29\pm0.13 $&$ -3.22\pm0.06 $\cr
\noalign{\vskip 3pt}
\hglue 0em SPIDER& $-58^{\circ}$& $ 236^{\circ}$& $ -3.34\pm0.06 $&$ -3.21\pm0.03 $\cr
\noalign{\vskip 5pt\hrule\vskip 2pt}
}}
\endPlancktablewide 
\endgroup
\end{table}

\begin{figure}
  \begin{center}
    \includegraphics[width=\linewidth]{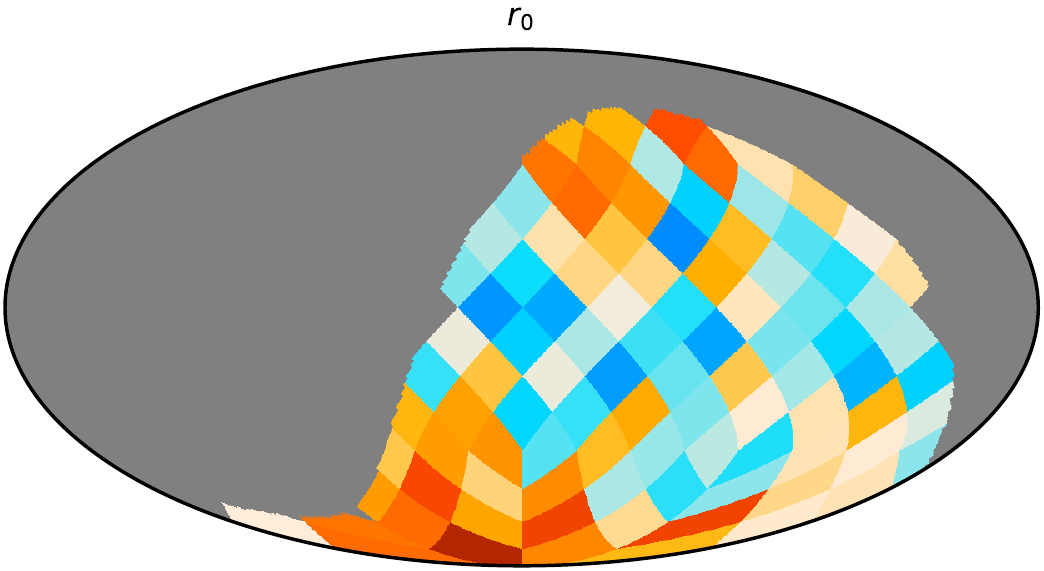}
    \includegraphics[width=\linewidth]{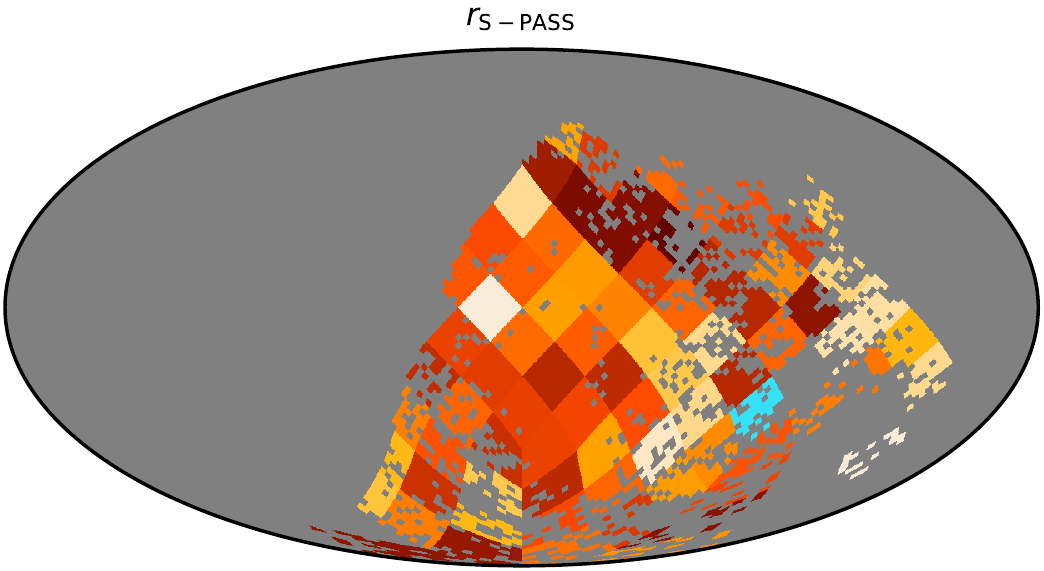}
    \includegraphics[width=\linewidth]{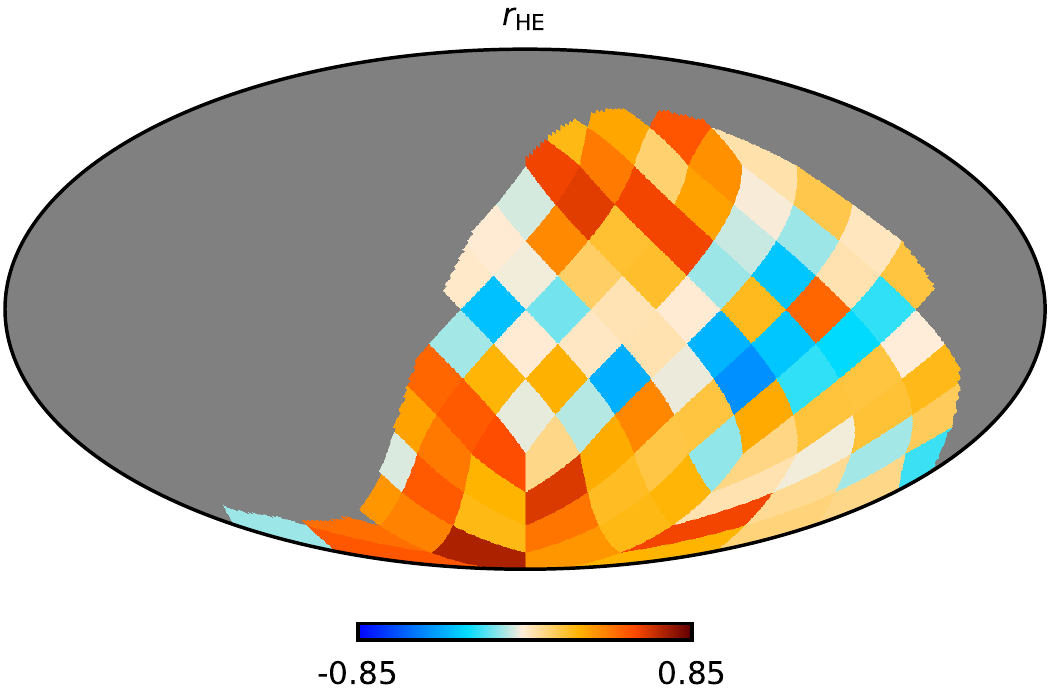}
  \end{center}
  \caption{Pearson correlation coefficient, $r$, evaluated between S-PASS data and \WMAP\ data in regions spanning $15^{\circ}\times15^{\circ}$. The \WMAP\ data are the same in all panels, while the S-PASS data are, from top to bottom, 1) the uncorrected data RM-0; 2) Faraday-corrected using RM-SPASS map \citep{Carretti:2019}; and 3) Faraday-corrected using RM-HE \citep{Hutschenreuter:2019}.}
  \label{fig:pearson}
\end{figure}

\begin{figure}
  \begin{center}
    \includegraphics[width=\linewidth]{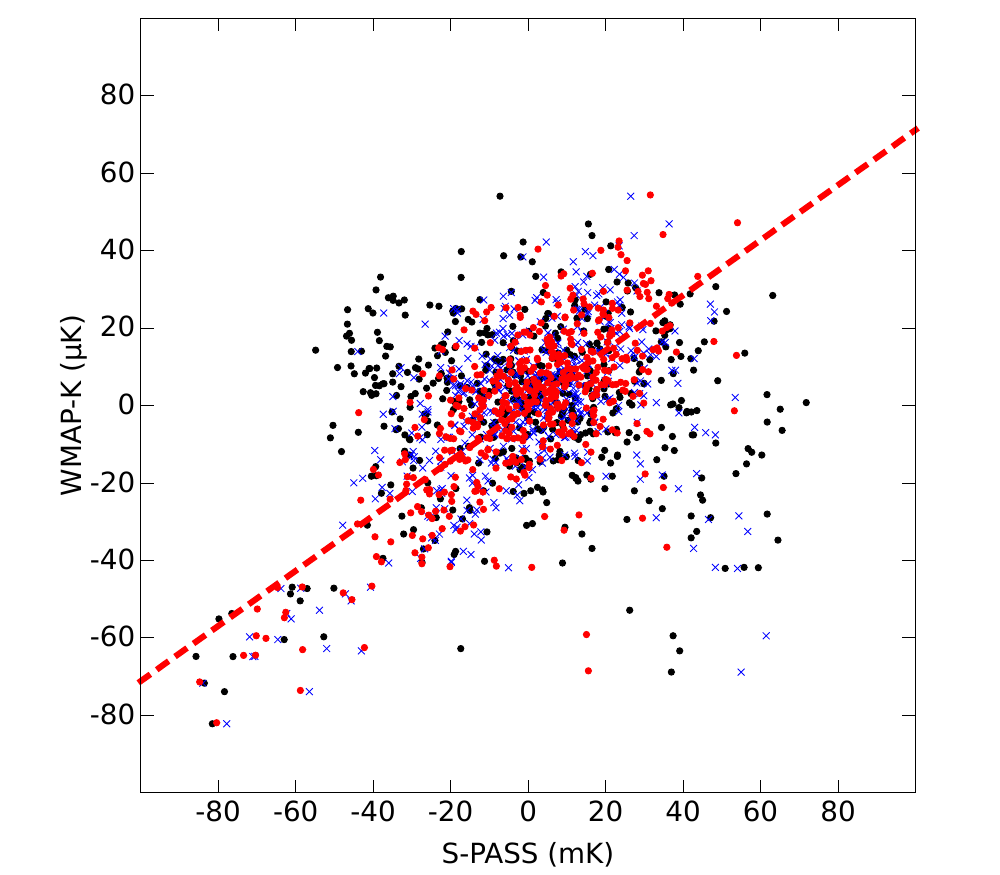}
  \end{center}
  \caption{Comparison of $T$--$T$ scatter plots for region 21 using different Faraday correction models. Both Stokes $Q$ and $U$ parameters are included in this figure, and the various symbols show 1) the uncorrected RM-0 data (black); 2) the Faraday-corrected data using RM-SPASS \citep{Carretti:2019} (red), and 3) the Faraday-corrected data using RM-HE \citet{Hutschenreuter:2019} (blue, crosses). The red dashed line is the best-fit slope using RM-SPASS.}
  \label{fig:pearson_scatter}
\end{figure}

First, in order to trace spatial variations in both the synchrotron
spectral index and the quality of the Faraday rotation model, we
divide the full S-PASS sky field according to $N_{\textrm{side}}=4$
regions, such that each region covers roughly
$15^{\circ}\times15^{\circ}$ and contains 256 $N_{\textrm{side}}=64$ pixels. 
Along the edge of the S-PASS survey,
some of these regions are only partially filled, and we exclude any
region for which more than half of the 256 pixels are excluded by
the survey geometry.  A total of 95 regions are retained by this
criterion, as shown in Fig.~\ref{fig:regions}. For the RM-SPASS maps, 
there are many missing pixels, and we have chosen to exclude regions 
where more than 75\% of the pixels are missing, 
in this case ending up with 79 regions.
Precise center locations for each region are listed in Table~\ref{tab:betas}.

Second, for each region $i$, we compute the Pearson correlation
coefficient between the Faraday-corrected S-PASS and the \WMAP\ sky
maps, minimized over local coordinate system orientations $\alpha$,
\begin{equation}
r_i = \min_{\alpha}\frac{\sum_{p\in i}
    Q_{p,\alpha}^{\mathrm{WMAP}}\hat{Q}_{p,\alpha}^{\mathrm{S-PASS}}}{\sqrt{\sum_{p
    \in i}
      Q_{p,\alpha}^{\mathrm{WMAP}}Q_{p,\alpha}^{\mathrm{WMAP}} \times
      \sum_{p \in i}
      \hat{Q}_{p,\alpha}^{\mathrm{S-PASS}}\hat{Q}_{p,\alpha}^{\mathrm{S-PASS}}}},
\label{eq:corrcoeff}
\end{equation}
where
\begin{equation}
  Q_{p,\alpha} = Q_p\,\cos 2\alpha + U_p\,\sin 2\alpha
  \label{eq:rotation}
\end{equation}
is the Stokes $Q$ parameter for pixel $p$ measured in a coordinate 
system that is rotated by an angle $\alpha$ relative to the reference 
system, and is normalized by subtracting the average value. 
We note that $\alpha=0^{\circ}$ corresponds to the un-rotated Stokes $Q$
parameter, while $\alpha=45^{\circ}$ corresponds to Stokes $U$. The
motivation for performing this minimization procedure is simply to
ensure that $r$ is measured in the coordinate system with the lowest
correlation (so that the reported $r$ is a worst-case scenario). 
These correlation values should be interpreted with some caution. 
The numbers are not the true measure of the correlation in a region 
since we are only reporting the lowest value. We are only using them 
to compare the different data sets, and to exclude regions with obvious 
low correlation. We will also perform a similar coordinate
system rotation when estimating the spectral index of polarized
synchrotron emission, as was also done in \citet{Fuskeland:2014eoa}.

Sky maps of $r$ are plotted in Fig.~\ref{fig:pearson} for each of the
three models; RM-0 (top panel), RM-SPASS (middle panel), and RM-HE
(bottom panel). The mean correlation coefficients averaged across the
sky are $r_{0} = 0.04\pm0.3$, $r_{\mathrm{S-PASS}} = 0.46\pm0.2$,
  and $r_{\mathrm{HE}}=0.16\pm0.3$, respectively. Thus, while both
RM-SPASS and RM-HE improves the overall correlation between S-PASS and
\WMAP, it is clear that the former yields an overall tighter agreement
between the two data sets. This is of course not unexpected,
considering the very different approaches taken by the two algorithms,
in particular recognizing the fact that RM-SPASS exploits \WMAP\ data
directly, while RM-HE does not. Also, as mentioned previously, RM-HE 
is measured using extra-galactic point sources, and might not be very 
well suited for this analysis of diffuse emission.

As a direct visualization of the corrections introduced by each of
these models, Fig.~\ref{fig:pearson_scatter} shows $T$--$T$ scatter
plots between the S-PASS and \WMAP\ Stokes $Q$ and $U$
parameters for region 21. Clearly, the correlation is visually tighter
for RM-SPASS than for either of the other two models, in agreement with
the quantitative results reported above.

Returning for a moment to Fig.~\ref{fig:spass}, the third row shows
the S-PASS sky map after Faraday correction with the RM-SPASS model,
while the bottom row shows the difference between the uncorrected and
corrected S-PASS maps. Comparing the first and third rows, we see that
the agreement with \WMAP\ significantly improves after applying the
Faraday correction. Furthermore, comparing the two bottom panels, we
note that the magnitude of the Faraday correction ranges between a few
percent to a factor of several tens. It is non-negligible in most areas
on the sky, and is therefore essential to take into account
in any joint analysis that combines S-PASS data with other
observations.  Based on these findings, we adopt the RM-SPASS model in
the following.

\begin{figure*}
  \begin{center}
    \includegraphics[width=\linewidth]{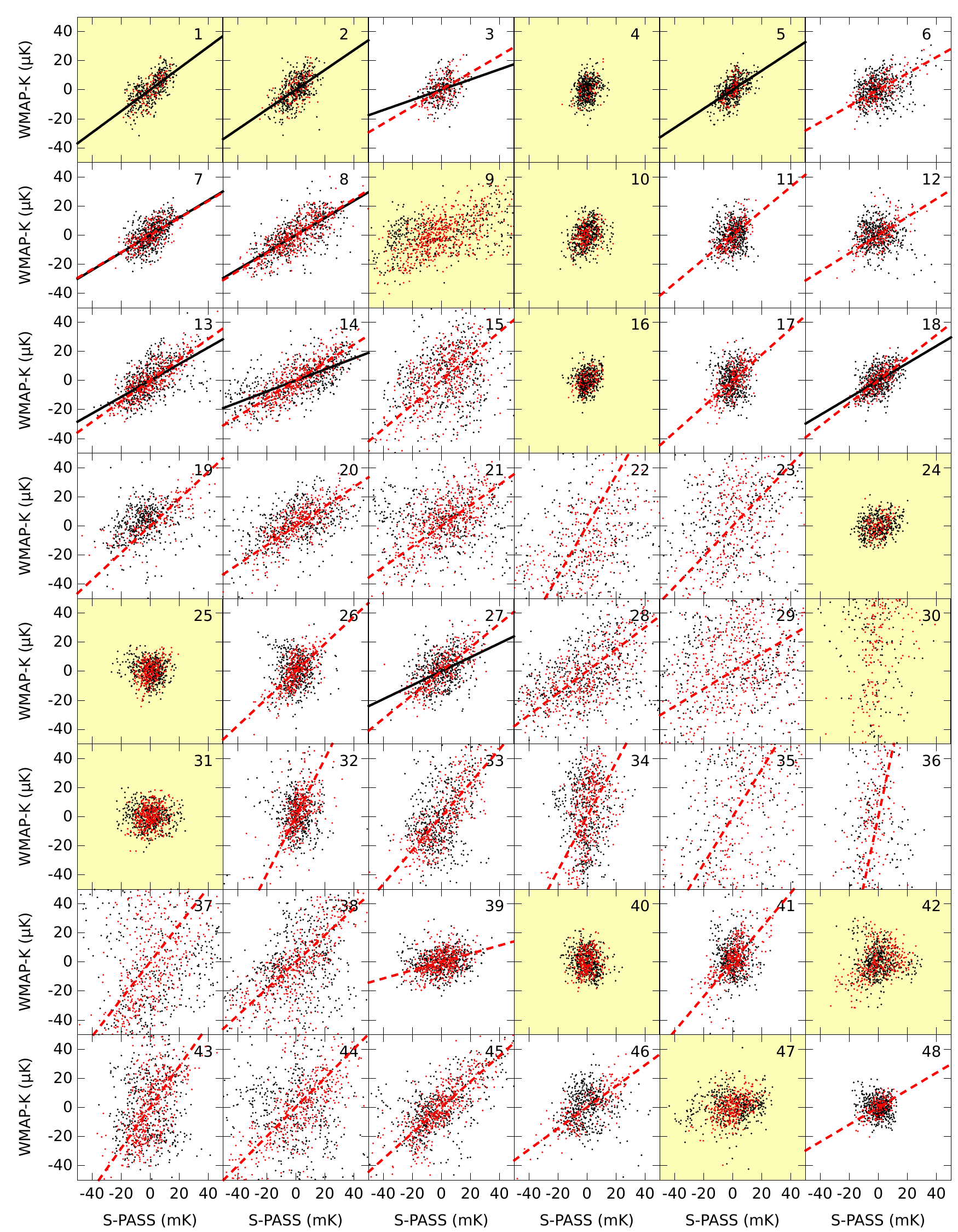}
  \end{center}
  \caption{$T$--$T$ plots for Stokes $Q$ and $U$ for the uncorrected data (black, solid) and for the Faraday-corrected data using RM-SPASS (red, dashed). The lines are the fitted values of $\beta_{\textrm{tot}}$. The yellow plots are where $r_{\mathrm{SPASS}}<0.2$ or $N_{\mathrm{pix}}<64$. Regions 1-48.}
  \label{fig:scatterplots1}
\end{figure*}

\begin{figure*}
  \begin{center}
    \includegraphics[width=\linewidth]{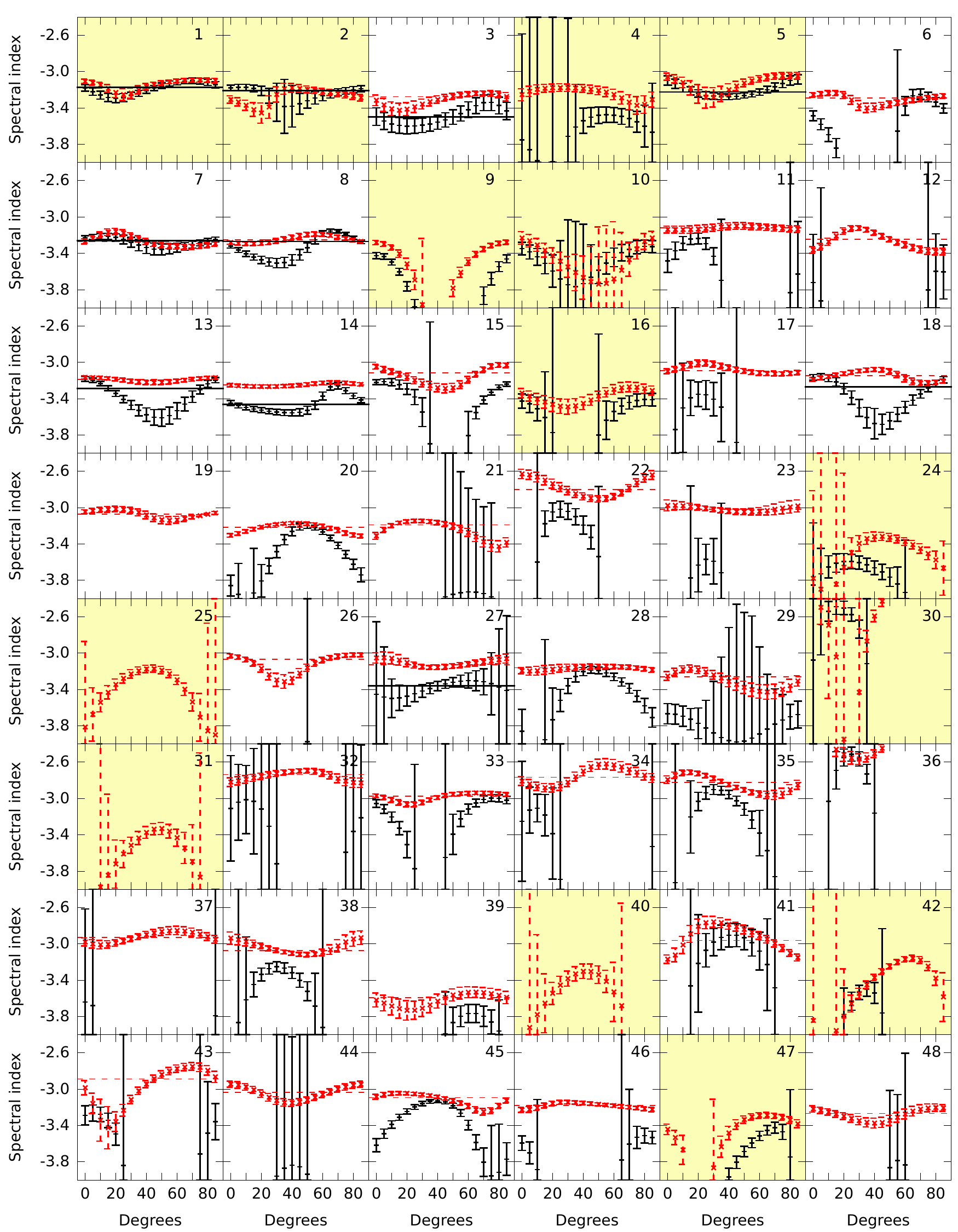}
  \end{center}
  \caption{Spectral index as a function of rotation angle for the uncorrected data (black, solid) and for the Faraday-corrected data (red, dashed). The lines are the values of $\beta_{\textrm{tot}}$. The yellow plots are where $r_{\mathrm{SPASS}}<0.2$, or $N_{\mathrm{pix}}<64$. Regions 1-48.}
  \label{fig:alphaplots1}
\end{figure*}

\begin{figure*}
  \begin{center}
    \includegraphics[width=\linewidth]{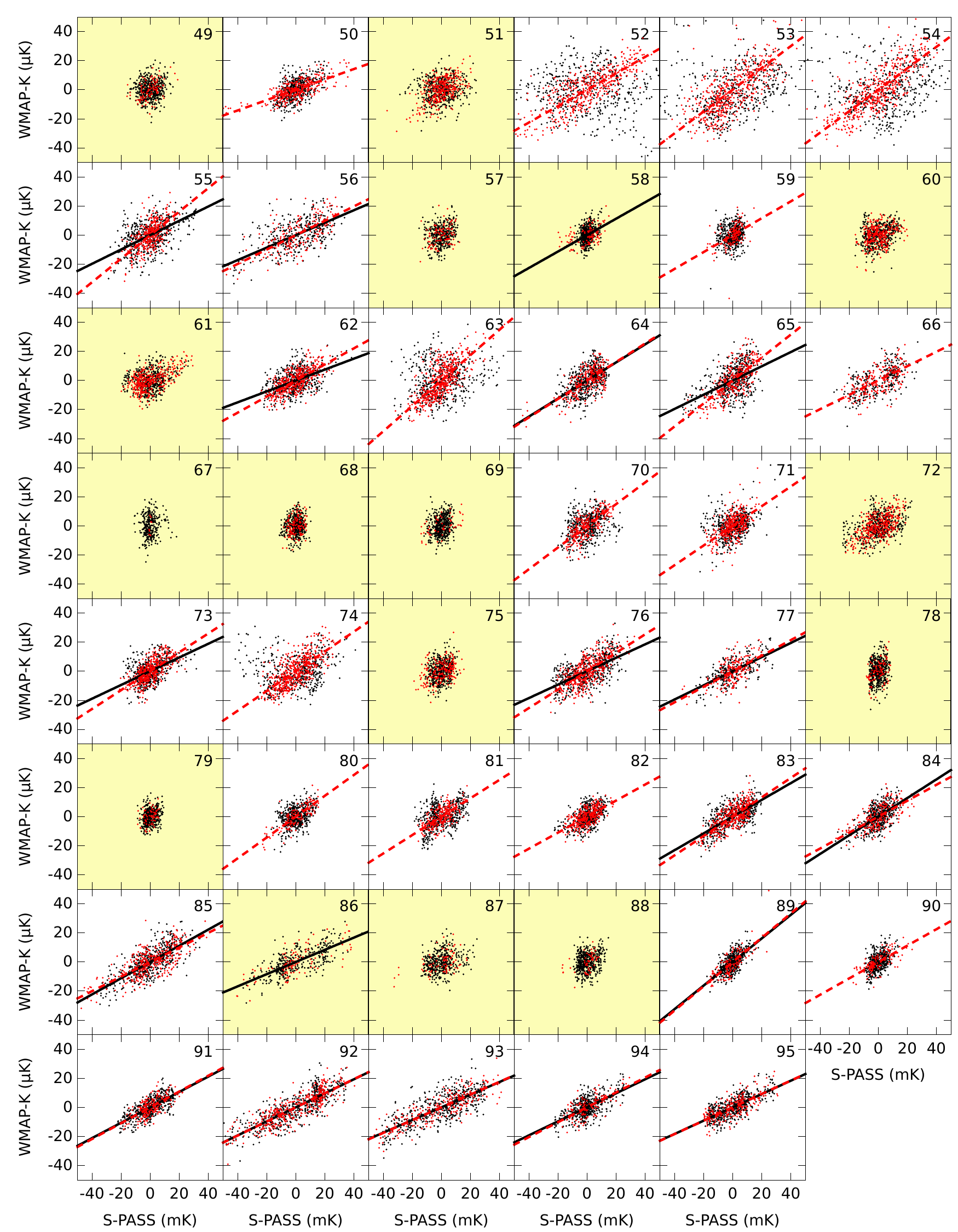}
  \end{center}
  \caption{$T$--$T$ plots for Stokes $Q$ and $U$ for the uncorrected data (black, solid) and for the Faraday-corrected data using RM-SPASS (red, dashed). The lines are the fitted value of $\beta_{\textrm{tot}}$. The yellow plots are where $r_{\mathrm{SPASS}}<0.2$ or $N_{\mathrm{pix}}<64$. Regions 49-95.}
  \label{fig:scatterplots2}
\end{figure*}

\begin{figure*}
  \begin{center}
    \includegraphics[width=\linewidth]{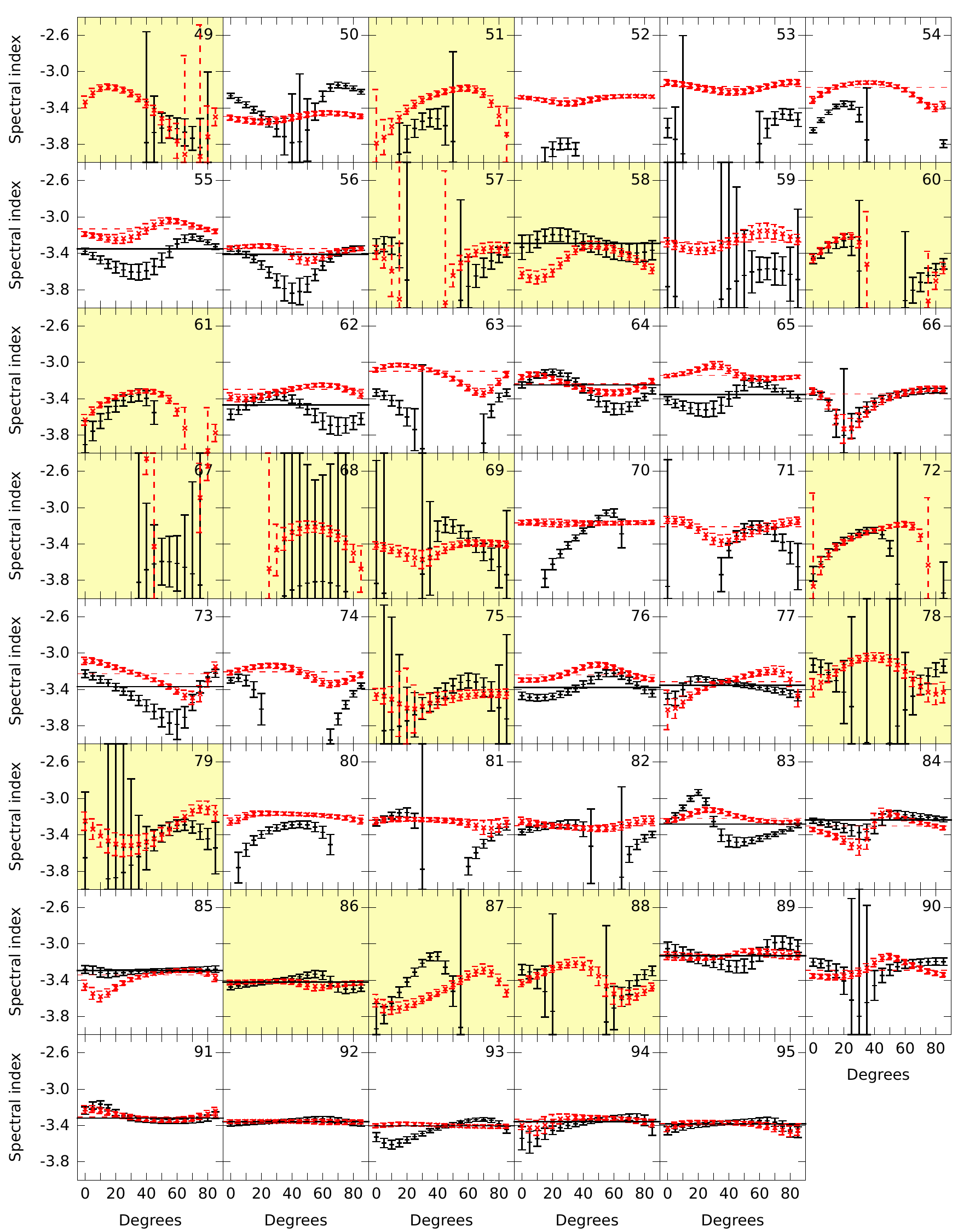}
  \end{center}
  \caption{Spectral index as a function of rotation angle for the uncorrected data (black, solid) and for the Faraday-corrected data (red, dashed). The lines are the values of $\beta_{\textrm{tot}}$. The yellow plots are where $r_{\mathrm{SPASS}}<0.2$ or $N_{\mathrm{pix}}<64$. Regions 49-95.}
  \label{fig:alphaplots2}
\end{figure*}

\section{Constraints on the spectral index of polarized synchrotron emission}
\label{sec:methods}

\subsection{Formalism}
\label{sec:formalism}

Our main goal in this paper is to use the S-PASS observations to
constrain the spectral index of polarized synchrotron emission,
$\beta_{\mathrm{s}}$. This parameter is defined by assuming that the
effective spectral energy density of synchrotron emission follows a
straight power law over the frequencies of interest. That is, we
assume that the observed data, $\mathbf{d}$, can be modeled as
\begin{equation}
\mathbf{d}_{\nu}(p) = \mathbf{A}(p) \left(\frac{\nu}{\nu_0}\right)^{\beta_i} + \mathbf{n}_{\nu}(p),
\label{eq:model0}
\end{equation}
where $\mathbf{d}_{\nu}(p)$ denotes a vector of the Stokes $Q$ and $U$
parameters at frequency $\nu$ in pixel $p$; $\mathbf{A}$ is the
amplitude of the signal at some reference frequency $\nu_0$; $\beta_i$
is the spectral index in region $i$; and $\mathbf{n}_{\nu}$ denotes
instrumental noise, which is typically assumed Gaussian with zero mean
and known (co-)variance.

For such a simple model, one of the most robust standard methods for
estimating $\beta_{\mathrm{s}}$ is through so-called $T$--$T$
plots. \citet{Fuskeland:2014eoa} applied this method to the \WMAP\ 23
and 33~GHz data, instead of the \WMAP\ 23~GHz and S-PASS 2.3~GHz data
as we do here. We therefore refer the interested reader to that paper
for full algorithmic details, and summarize only the main points here.

In the special case of noiseless data ($\mathbf{n}=0$), we see from Eq.~\ref{eq:model0}
that the spectral index $\beta_{\mathrm{s}}$ may be estimated from only two
different data points through the following ratio,
\begin{equation}
\frac{d_{\nu_1}}{d_{\nu_2}} =
\left(\frac{\nu_{1}}{\nu_2}\right)^{\beta_{\mathrm{s}}} \implies \beta_{\mathrm{s}} = 
\frac{\ln(d_{\nu_1}/d_{\nu_2})}{\ln (\nu_1/\nu_2)}.
\end{equation}
Analogously, for noisy data one may fit a straight line, $y = ax+b$, to the
distribution of pairs of observation, $\{d_{\nu_1}(p),d_{\nu_2}(p)\}$,
and compute $\beta_{\mathrm{s}}$ via the slope of the fitted line,
\begin{equation}
d_{\nu_1}(p) = a \, d_{\nu_2}(p) + b = \left(\frac{\nu_{1}}{\nu_2}\right)^{\beta_{\mathrm{s}}}
d_{\nu_2}(p) + b 
\implies \beta_{\mathrm{s}} = \frac{\ln a}{\ln (\nu_1/\nu_2)}.
\end{equation}
This is called the $T$--$T$ plot technique, and it is a widely used tool
in radio astronomy. The only slightly subtle point in this procedure
is how to fit the straight line in the presence of noise in both data
sets. However, several algorithms have been developed for precisely
this purpose, and we adopt the effective variance method of
\citet{orear1982}, as implemented and described by
\citet{Fuskeland:2014eoa}.

To obtain robust results that are independent of the orientation of
the (Galactic) coordinate system used to pixelize the S-PASS and \WMAP\
data, we marginalize over polarization angle, similar to the procedure
adopted for Faraday rotation assessment in the previous section. That
is, we rotate the original data sets by an angle $\alpha$ into a new
coordinate system by Eq.~\ref{eq:rotation}, considering all values of
$\alpha$ between $0^\circ$ and $85^\circ$ in steps of $5^\circ$. We
then estimate $\beta_{\mathrm{s}}$ using the $T$--$T$ plot approach for each value
of $\alpha$, and report either the full function $\beta_{\mathrm{s}}(\alpha)$ or
the corresponding inverse-variance weighted mean
\begin{equation}
\beta_{\textrm{tot}} = \frac{\sum_{i=1}^{18} \beta_i / \sigma_i^2}{\sum_{i=1}^{18} 1 / \sigma_i^2},
\end{equation}
where $\sigma_i$ is the uncertainty for a given value of $\alpha$; see
Eq.~14 in \citet{Fuskeland:2014eoa}. These uncertainties are estimated
by adding the statistical and systematic uncertainties in quadrature,
and the systematic uncertainty is estimated using bootstrap
sampling. That is, we randomly draw 10\,000 new data combinations from
the original data set, allowing duplicate points. Then the spectral
indices are calculated for each new data set, and the standard
deviation of this distribution is adopted as a systematic
uncertainty.

%%%%%%%%%%%%%%%%%%%%%%%%%%%%%%%%%%%%%
\subsection{Results}
\label{sec:results}

We now apply the method outlined above for each of the 95 regions
defined in Fig.~\ref{fig:regions} to the Faraday-corrected S-PASS
2.3~GHz and the \WMAP\ 23~GHz sky maps. Figure~\ref{fig:scatterplots1}
shows individual $T$--$T$ scatter plots for regions 1 through 48, for 
the Stokes $Q$ and $U$ parameters. The uncorrected and the 
RM-SPASS Faraday-corrected data are shown as black and red points,
respectively. The best-fit straight lines when evaluating $\beta_i$ 
for all rotation angles $\alpha$ and taking the inverse-variance 
weighted means, are shown as their respective solid black and dashed red lines. 
Regions for which Pearson's correlation coefficient is smaller than 0.2 are
excluded from the analysis, and no best-fit lines are indicated in
these cases. Also excluded are regions using the RM-SPASS data where
more than 75\% of the pixels are missing ($N_{\mathrm{pix}}<64$). This results
in a few regions where only the uncorrected data are used 
(regions $1$, $2$, $5$, $58$ and $86$).
Both of these types of excluded regions are flagged with a yellow background in
Figs.~\ref{fig:scatterplots1}-~\ref{fig:alphaplots2}.

We observe large variations between the different regions in this figure.
For instance, while regions 8 and 13 exhibit visually obvious
correlations between S-PASS and \WMAP, others, such as regions 12 and 48,
require detailed statistical analysis to pick out a statistically
significant correlation. Some regions show a large overall scatter,
indicating that there are large signal variations inside these regions,
while others show very small scatter and are dominated by instrumental
noise. 

Using the $T$--$T$ plot method, we worked under the assumption that there is 
a common spectral index for all pairs of pixels within a region. However, this 
may not always be the case for our regions. So some of the scatter may be 
because of internal variation of the spectral index in a region.

Figure~\ref{fig:alphaplots1} shows the corresponding constraints on
the spectral index $\beta_i$ as a function of rotation angle $\alpha$
for the same set of regions. Solid black points show results for
uncorrected S-PASS data, while dashed red points show results for the
RM-SPASS Faraday-corrected data. The horizontal lines indicate the
respective inverse-variance weighted means.

\begin{figure}
  \begin{center}
    \includegraphics[width=\linewidth]{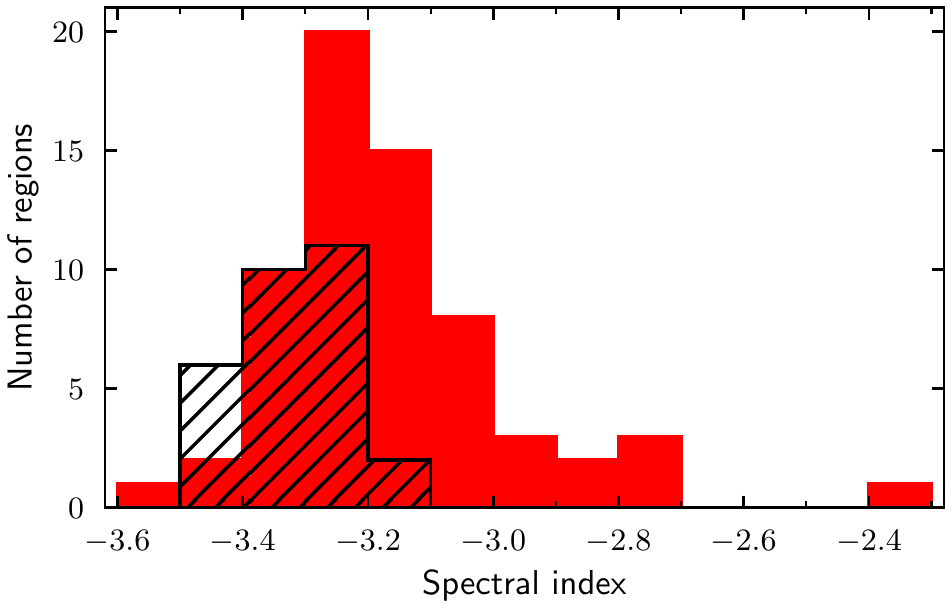}
  \end{center}
  \caption{Histogram of the polarized synchrotron spectral indices using the $T$--$T$ plot method on S-PASS 2.3~GHz and \WMAP\ 23~GHz data for the 19 accepted regions using uncorrected S-PASS data (black) and for the 65 accepted regions using Faraday-corrected S-PASS data (red).}
  \label{fig:hist_betas}
\end{figure}

Cases for which the black and red points agree closely primarily
correspond to regions in which the magnitude and impact of the Faraday
correction model is modest. This happens most typically at high
Galactic latitudes, where the Galactic magnetic field is weak. The
most typical case, however, is that the red points exhibit better
coherence than the black points, suggesting that the Faraday
correction is both significant and beneficial. 
Corresponding results for regions 49--95 are shown in
Figs.~\ref{fig:scatterplots2} and \ref{fig:alphaplots2}.

\begin{figure}
  \begin{center}
    \includegraphics[width=\linewidth]{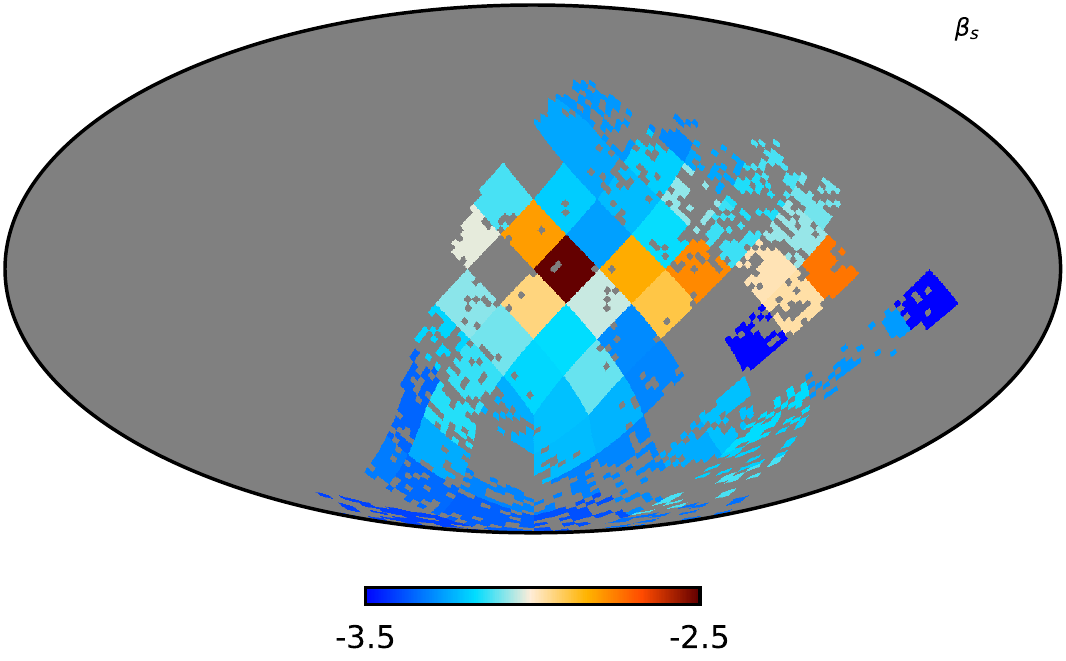}
  \end{center}
  \caption{Spatial distribution of the polarized synchrotron spectral index between the Faraday-corrected S-PASS 2.3~GHz and \WMAP\ 23~GHz data. Only regions for which the Pearson correlation coefficient $r>0.2$ are shown.}
  \label{fig:map_betas}
\end{figure}

\begin{figure}
  \begin{center}
    \includegraphics[width=\linewidth]{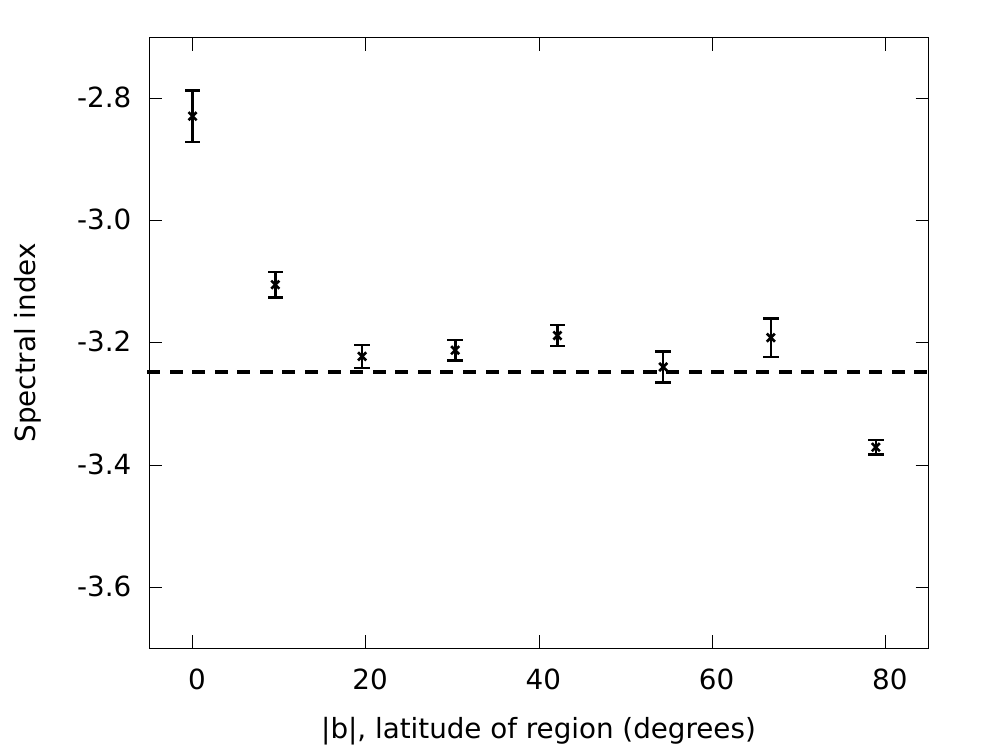}
  \end{center}
  \caption{
Polarized synchrotron spectral index as a function of the absolute value of latitude for the Faraday-corrected data (RM-SPASS). Regions at the same latitude are combined by estimating the inverse-variance weighted mean. Only regions for which the Pearson correlation coefficient is $r>0.2$ are included. The horizontal line shows the inverse-variance weighted mean values of all regions, $\beta_{\textrm{tot}}=-3.24$.}
  \label{fig:latitude_betas}
\end{figure}

Region 73 is an example of another interesting case. Here we observe
large drifts as a function of rotation angle, but with very small
uncertainty at each individual angle. Statistically speaking, there is
a 3--$4\sigma$ discrepancy between the spectral indices observed at
$\alpha=10^{\circ}$ and $70^{\circ}$, with values ranging between
$-3.1$ and $-3.5$. Taken at face value, this could in principle
be interpreted as evidence for statistically significant variations in
the spectral indices of the two Stokes parameters, $Q$ and $U$, which
is entirely possible from a physical point of view: Local alignment
with the Galactic magnetic field or true spatial variations along each
line-of-sight are only two physical effects that could create such a
signal. However, very large variations are difficult to interpret in
terms of physical variations in the local electron energy
distribution. The applied RM maps may 
also rotate the low-frequency signal both in or out of phase with the 
high-frequency signal, resulting in either too shallow or too steep spectral index.

To account for further systematical uncertainties in the analysis, we conservatively adopted
\begin{equation}
\sigma_{\beta_{\mathrm{s}}} \equiv (\max_\alpha \beta_{\mathrm{s}}(\alpha) - \min_\alpha \beta_{\mathrm{s}}(\alpha))/2
\end{equation}
as our final systematic estimate of the uncertainty on $\beta_{\mathrm{s}}$,
evaluated separately for each region. This 
is added in quadrature to the uncertainty
defined by the inverse-variance weighted average, which takes into
account both the statistical uncertainty and the systematic 
uncertainty from the bootstrap procedure explained in 
Sect.~\ref{sec:formalism}. The statistical uncertainty gives 
a negligible contribution compared to the two systematical ones. 
All reported values are using the total uncertainty.

To test the impact of the errors that are distributed together with the RM maps from S-PASS, a full analysis have been made on S-PASS data that have been coherently rotated by $\phi_{\mathrm{RM-SPASS}} + 1\sigma_{\mathrm{RM-SPASS}}$. This results in only a small shift of the regional spectral indices; maximum $1 \sigma$ deviation in three regions and usually significantly less.

Final spectral index estimates for each region with correlation coefficients
higher than 0.2, as defined by Eq.~\ref{eq:corrcoeff} and shown in
Fig.~\ref{fig:pearson}, are listed in Table~\ref{tab:betas}. Without
Faraday correction, this includes 29 regions, while with
RM-SPASS-based Faraday correction a total of 65 regions exceed the cut
criterion. Figure~\ref{fig:hist_betas} shows a histogram of these values.

Inverse-variance weighting of the estimates for all regions
yields a full-sky average of $\beta_{\mathrm{s}}=-3.24\pm0.01$ with Faraday
corrections and $\beta_{\mathrm{s}}=-3.32\pm0.02$ without Faraday correction. 
The corresponding standard deviations are $0.09$ and $0.19$, respectively.
These results are in excellent agreement with constraints
derived from S-PASS and \WMAP\ by \citet{krachmalnicoff:2019} using
power spectra as their primary tool, reporting a full-sky average
spectral index for polarized synchrotron emission of
$\beta_{\mathrm{s}}=-3.22\pm0.08$.

\begin{figure*}
  \begin{center}
    \includegraphics[width=0.49\linewidth]{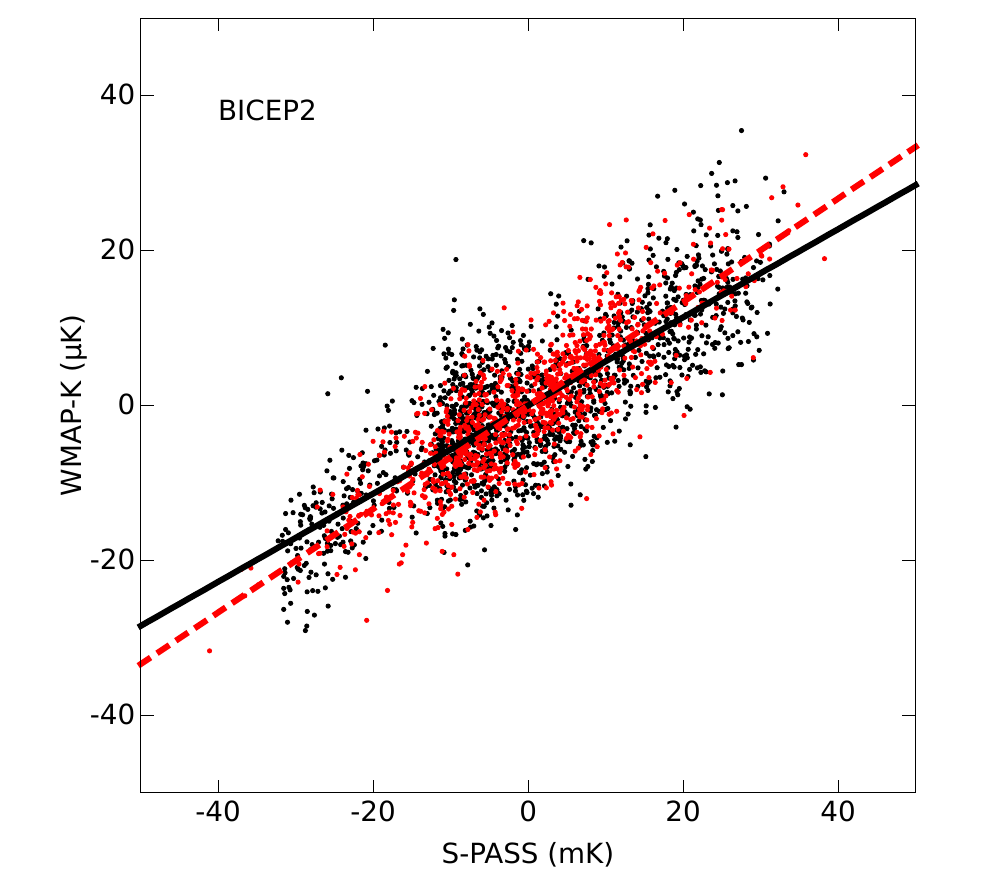}
    \includegraphics[width=0.49\linewidth]{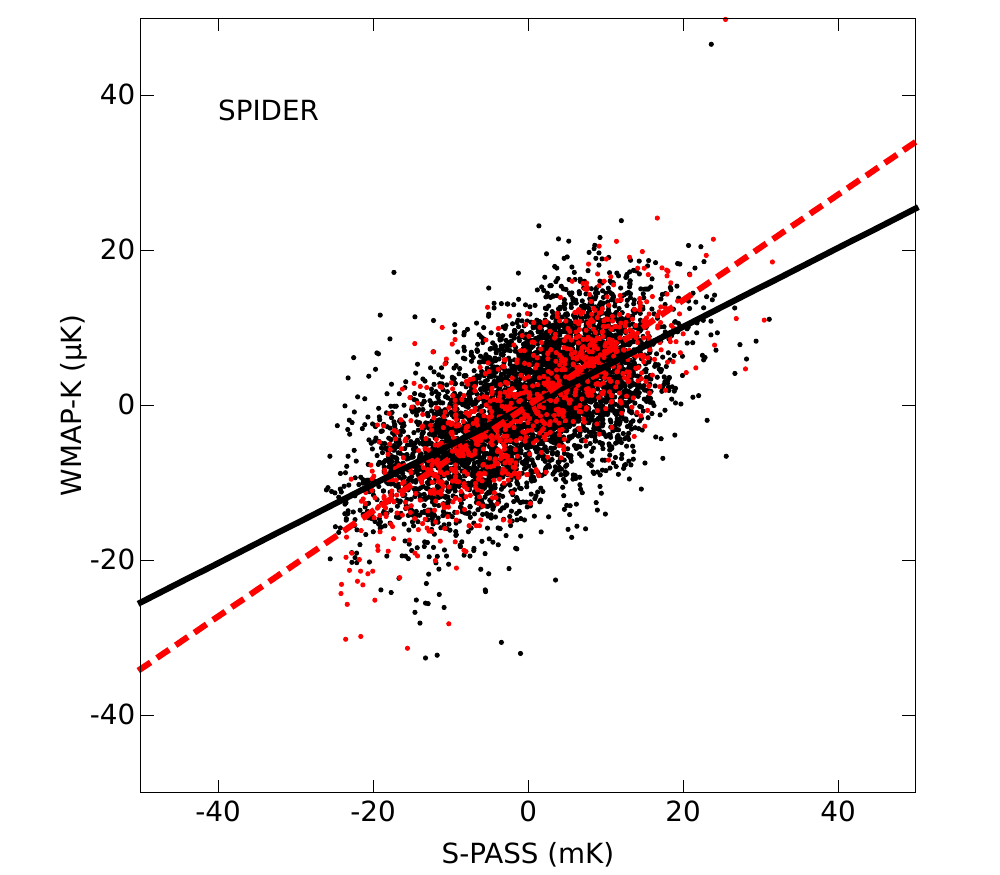}\\
    \includegraphics[width=0.49\linewidth]{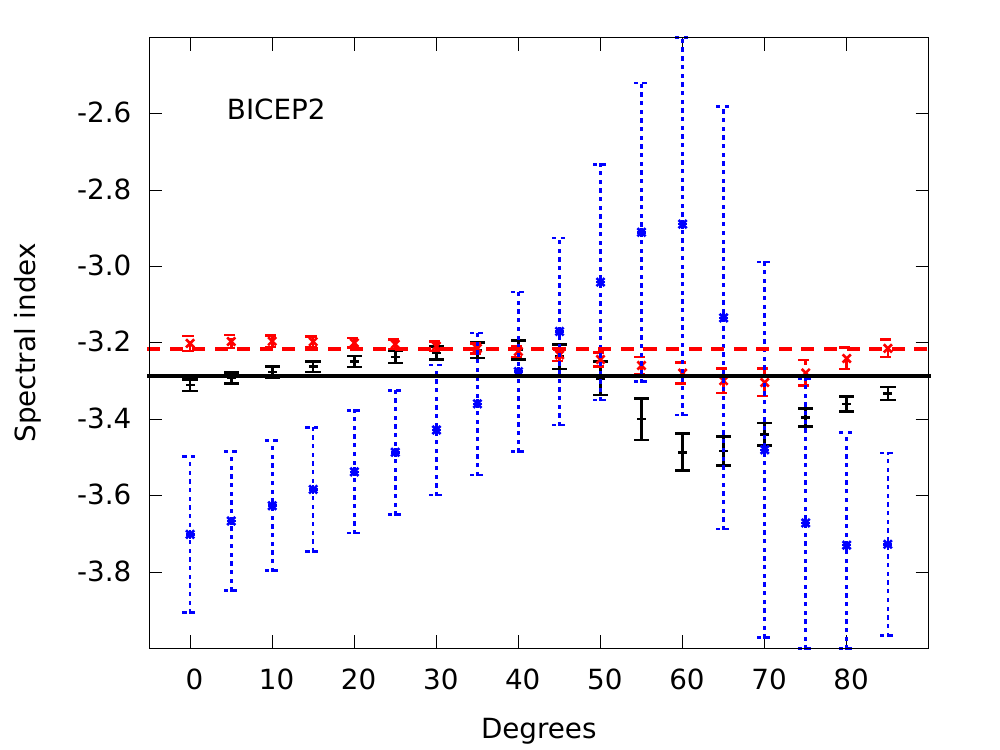}
    \includegraphics[width=0.49\linewidth]{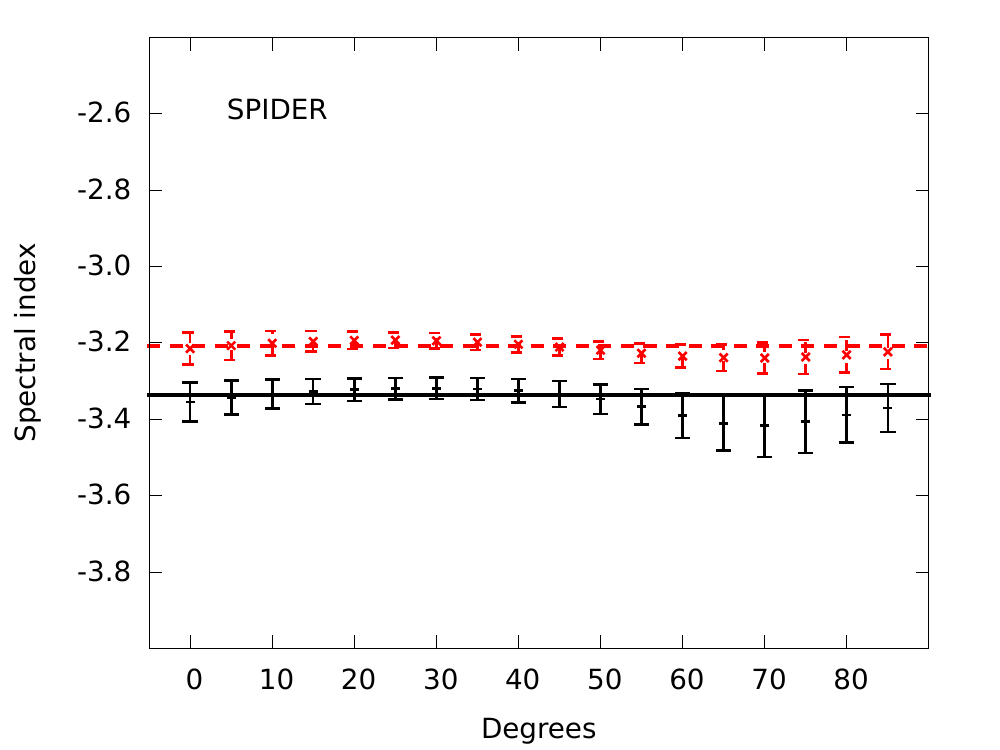}
  \end{center}
  \caption{Results for the BICEP2 (\emph{left}) and SPIDER (\emph{right}) regions. The blue values are from Fig.~9 in \citet{Fuskeland:2014eoa} using the $T$--$T$ plot method between the \WMAP\ 23~GHz and 33~GHz data.}
  \label{fig:bicep_scatter_alpha}
\end{figure*}

Figure~\ref{fig:map_betas} shows the spatial distribution of the mean
spectral index of polarized synchrotron emission for each accepted
region. Here we clearly see a statistically significant and systematic
spatial variation in $\beta_{\mathrm{tot}}$, in the form of index steepening 
from low to high galactic latitudes. To quantify this observation further, we
plot in Fig.~\ref{fig:latitude_betas} the average spectral index as a
function of the absolute value of the Galactic latitude, $|b|$. Within
each latitude bin, the various spectral indices have been
inverse-variance weighted to produce a joint estimate, adopting the
same methodology as described above for the full-sky average. Based on
these measurements, we find that the spectral index typically range
between $\beta_{\mathrm{s}}=-2.7$ and $-2.9$ along the Galactic plane, and between
$\beta_{\mathrm{s}}=-3.1$ and $-3.3$ at high Galactic latitudes.  
The flat spectral index at low Galactic latitude are likely partly due to 
depolarization effects, which reduce the S-PASS amplitude and thus reduce 
the spectral index. This effect is only dominant in proximity of the Galactic 
plane, whereas for instance \citet{krachmalnicoff:2019} has chosen to exclude 
that region.
Qualitatively speaking, this general behavior of the spectral index is 
in good agreement with the conclusions of numerous previous analyses, 
including \citet{kogut2007,Fuskeland:2014eoa,krachmalnicoff:2019}, all 
reporting significant steepening from low to high Galactic latitudes.

%%%%%%%%%%%%%%%%%%%%%%%%%%%%%%%%%%%%%
\subsection{Polarized synchrotron emission in the BICEP2 and SPIDER fields}
\label{sec:special_cases}

Next, we consider two special cases of particular
interest with respect to current and upcoming constraints on the
tensor-to-scalar ratio, namely those corresponding to the BICEP2
\citep{bicep2} and SPIDER \citep{nagy:2017} fields. The BICEP2 field
is defined approximately by a rectangle in celestial coordinates
spanning $-40^{\circ} < \mathrm{RA} < 40^{\circ}$, $-65^{\circ} <
\mathrm{dec} < -50^{\circ}$, and covers about 1\,\% of the sky. The
central part of the SPIDER field is defined by $30^{\circ} <
\mathrm{RA} < 70^{\circ}$, $-55^{\circ} < \mathrm{dec} < -15^{\circ}$,
and covers about 8\,\% of the sky.
The two fields are analyzed in the same way as the previous regions,
the number of pixels being 2305 (SPIDER) and 773 (BICEP2) in the 
uncorrected data, and reduced to 496 (SPIDER) and 486 (BICEP2) 
for the Faraday-corrected data due to the missing pixels.

The top panels of Fig.~\ref{fig:bicep_scatter_alpha} show $T$--$T$
plots between both the Faraday-corrected (red) and uncorrected (black) 
S-PASS and the WMAP data for each of these two fields. 
A strong correlation is observed in both cases. 
The bottom panels show corresponding $\beta_{\mathrm{s}}(\alpha)$ results
for each field, and here we see that $\beta_{\mathrm{s}}$ is well constrained for
any value of $\alpha$, suggesting that the final spectral index
estimates are robust with respect to both instrumental effects and
modeling errors. As reported in the bottom of Table~\ref{tab:betas},
the mean spectral indices are $\beta_{\mathrm{BICEP2}} = -3.22\pm0.06$
and $\beta_{\mathrm{SPIDER}} = -3.21\pm0.03$. For the
BICEP2 field, the blue points show similar constraints derived from the
WMAP 23 and 33~GHz data, as reported by \citet{Fuskeland:2014eoa};
these are in good agreement with the new estimates, only with a lower
signal-to-noise ratio. This previous analysis did not contain any 
analysis of the SPIDER field.

These estimates may be used to predict the absolute level of polarized
synchrotron emission at 90 and 150~GHz, the two primary CMB
frequencies for both BICEP2 and SPIDER, by extrapolating the observed
synchrotron amplitude at 23~GHz. To do this, we first computed two
independent \WMAP\ K-band ``half-mission'' maps by co-adding
\WMAP\ observation years 1--4 (HM1) and 5--9 (HM2). We
smoothed each map to an effective angular resolution of $2^{\circ}$ to
suppress uncorrelated noise. Next, we formed an unbiased estimate of the
square of the polarization amplitude per pixel by cross-correlating
the two half-mission maps,
\begin{equation}
  \hat{P}^2 = Q_{\mathrm{HM1}}Q_{\mathrm{HM2}} +
  U_{\mathrm{HM1}}U_{\mathrm{HM2}}.
\end{equation}
We note that because $\hat{P}^2$ is estimated as a cross-product between
two half-mission maps, it can take on negative values. However, this
can only happen due to $\hat{P}^2$ having anti-correlated noise fluctuations,
and not true signal variations. We therefore estimate the linear
polarization amplitude as
\begin{equation}
\hat{P} = \sqrt{\max(\hat{P}^2,0)},
\end{equation}
which is strictly positive. This quantity does not have a systematic
noise bias due to auto-correlations, but only from the positivity
prior, which is relevant only in low signal-to-noise regions.

\begin{figure}
  \begin{center}
    \includegraphics[width=\linewidth]{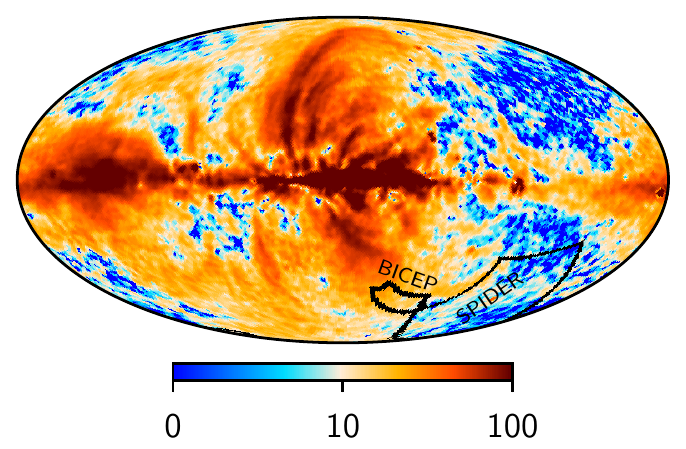}
    \includegraphics[width=\linewidth]{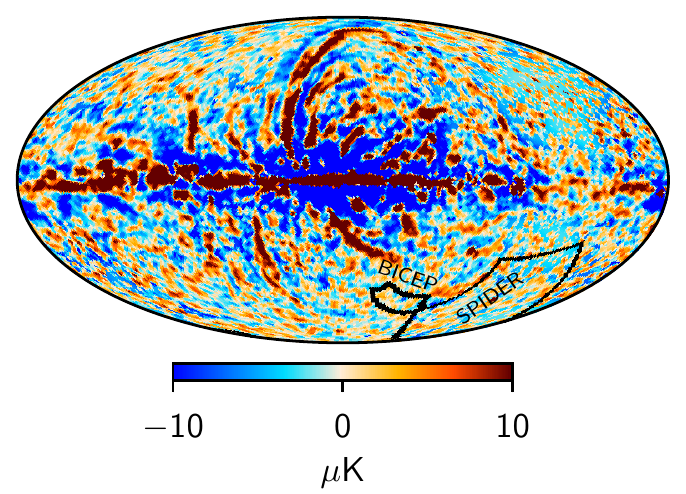}
  \end{center}
  \caption{\WMAP\ K-band polarization amplitude, $\hat{P}$, 
    estimated by cross-correlating two half-mission maps.
    The top panel is smoothed to an effective angular resolution of $2^{\circ}$
    FWHM. The bottom panel is after bandpass filtering
    to include only scales between $2^{\circ}$ and $10^{\circ}$,
    highlighting structures between $\ell\approx20$ and 100.}
  \label{fig:P}
\end{figure}

\begin{table}[t]  % Table 2  
\begingroup
\newdimen\tblskip \tblskip=5pt
\caption{Predictions for polarized synchrotron emission in the BICEP2
  and SPIDER fields at a smoothing scale of $2^{\circ}$ FWHM, based on
  \WMAP\ 23~GHz and S-PASS. The top section shows mean and standard
  deviation for the full map at $2^\circ$ FWHM smoothing scale, while
  the bottom section lists standard deviations for the difference
  between $\hat{P}$ smoothed to $2^{\circ}$ and $10^{\circ}$ FWHM. The
  mean for the latter is consistent with zero for both fields, due to
  instrumental noise in the \WMAP\ 23~GHz map.\label{tab:predictions}}
\vskip -6mm
%\scriptsize
\footnotesize
\setbox\tablebox=\vbox{
\newdimen\digitwidth
\setbox0=\hbox{\rm 0}
\digitwidth=\wd0
\catcode`*=\active
\def*{\kern\digitwidth}
\newdimen\signwidth
\setbox0=\hbox{+}
\signwidth=\wd0
\catcode`!=\active
\def!{\kern\signwidth}
\newdimen\decimalwidth
\setbox0=\hbox{.}
\decimalwidth=\wd0
\catcode`@=\active
\def@{\kern\signwidth}
\halign{\hbox to 0.9in{#\leaderfil}\tabskip=1.0em&
    \hfil#\hfil\tabskip=0.4em&
    \hfil#\hfil\tabskip=0em\cr
\noalign{\doubleline}
\omit\hfil Frequency (GHz) \hfil &$\hat{P}_{\mathrm{BICEP2}}$
($\mu\mathrm{K}$)&$\hat{P}_{\mathrm{SPIDER}}$ ($\mu\mathrm{K}$)\cr 
\noalign{\vskip 5pt\hrule\vskip 5pt}
\noalign{\vskip 3pt}
\multispan3\textit{Mean and RMS polarization amplitude of $\hat{P}(2^{\circ})$}\hfil\cr
\noalign{\vskip 2pt}      
*23 &$**18!**\pm6!****$&$7!**\pm5!***$\cr
*90 &$*0.25*\pm0.08*$&$0.10*\pm0.07*$\cr
150 &$*0.069\pm0.023$&$0.026\pm0.019$\cr
\noalign{\vskip 5pt}      
\multispan3\textit{RMS polarization amplitude difference of
  $\hat{P}(2^{\circ})-\hat{P}(10^{\circ})$}\hfil\cr
\noalign{\vskip 2pt}      
*23 &$\phantom{88.888\pm}\,<2.4**$&$\phantom{88.888\pm}\,<3.4**$\cr
*90 &$\phantom{88.888\pm}\,<0.03*$&$\phantom{88.888\pm}\,<0.05*$\cr
150 &$\phantom{88.888\pm}\,<0.009$&$\phantom{88.888\pm}\,<0.013$\cr
\noalign{\vskip 5pt\hrule\vskip 2pt}
}}
\endPlancktablewide 
\endgroup
\end{table}

The resulting \WMAP\ 23~GHz $\hat{P}$ map is shown in the top panel of
Fig.~\ref{fig:P}, with the BICEP2 and SPIDER regions indicated by
black lines. The bottom panel shows the same map, but after
subtracting itself smoothed to $10^{\circ}$ FWHM, thereby highlighting
multipole moments between $\ell\approx20$--100, or angular scales
between $2^{\circ}$ and $10^{\circ}$. The mean and standard deviation
of the full map is $\left<\hat{P}\right> = 18\pm6\,\mu\textrm{K}$
within the BICEP2 region, and $7\pm5\,\mu\textrm{K}$ within the
SPIDER region. Thus, the BICEP2 region has significantly higher
polarized synchrotron emission levels than the SPIDER field, but most
of this is only detectable on large angular scales. For the bandpass
filtered map, both regions have a mean consistent with zero, while the
standard deviations are $2.4\,\mu\textrm{K}$ for the BICEP2 region, and
$3.4\,\mu\textrm{K}$ for the SPIDER region. These values largely
reflects the instrumental noise level of the \WMAP\ 23~GHz map, and
they therefore only correspond to upper limits on the synchrotron
level in these fields, not a determination of the actual synchrotron
variation within each field.

We can now estimate the polarization amplitude at 90 and 150~GHz by
extrapolating $\hat{P}$ from \WMAP\ K-band (22.45~GHz), by scaling
according to a power law model in brightness temperature, while
properly accounting for unit conversions between brightness and
thermodynamic units. The total extrapolation factor is given by
\begin{equation}
  f(\nu) = \frac{g(\nu)}{g(22.45\,\mathrm{GHz})} \left(\frac{\nu}{22.45\,\mathrm{GHz}}\right)^{\beta_{\mathrm{s}}},
\end{equation}
where $g(\nu) = (e^x-1)^2/x^2e^x$, $x=h\nu/kT_{\mathrm{CMB}}$ is the
conversion factor between brightness temperature and thermodynamic
temperature units; $h$ and $k$ are the Planck and Boltzmann constants;
and $T_{\mathrm{CMB}}=2.7255\,\mathrm{K}$ is the CMB temperature
\citep{fixsen:2009}.

Table~\ref{tab:predictions} lists the extrapolated predictions for
each region and multipole range, based on the mean spectral indices
derived above. To set those values in context, we note that a
tensor-to-scalar ratio of $r=0.1$ induces a large-scale $B$-mode
signal with standard deviation equal to $0.08\,\mu\textrm{K}$ at a
smoothing scale of $2^{\circ}$ FWHM, while a ratio of $r=0.01$ induces
a $B$-mode standard deviation of $0.02\,\mu\textrm{K}$. After
high-pass filtering, the polarized synchrotron contribution is
therefore constrained to $r\lesssim0.02$ at 90~GHz and
$r\lesssim0.005$ at 150~GHz for either field, within a small factor
depending on the local noise properties of the \WMAP\ survey.

%%%%%%%%%%%%%%%%%%%%%%%%%%%%%%%%%%%%%
\subsection{Comparison with results in the 23-33 GHz range}
\label{sec:comparison_2014}

\begin{figure}
  \begin{center}
    \includegraphics[width=\linewidth]{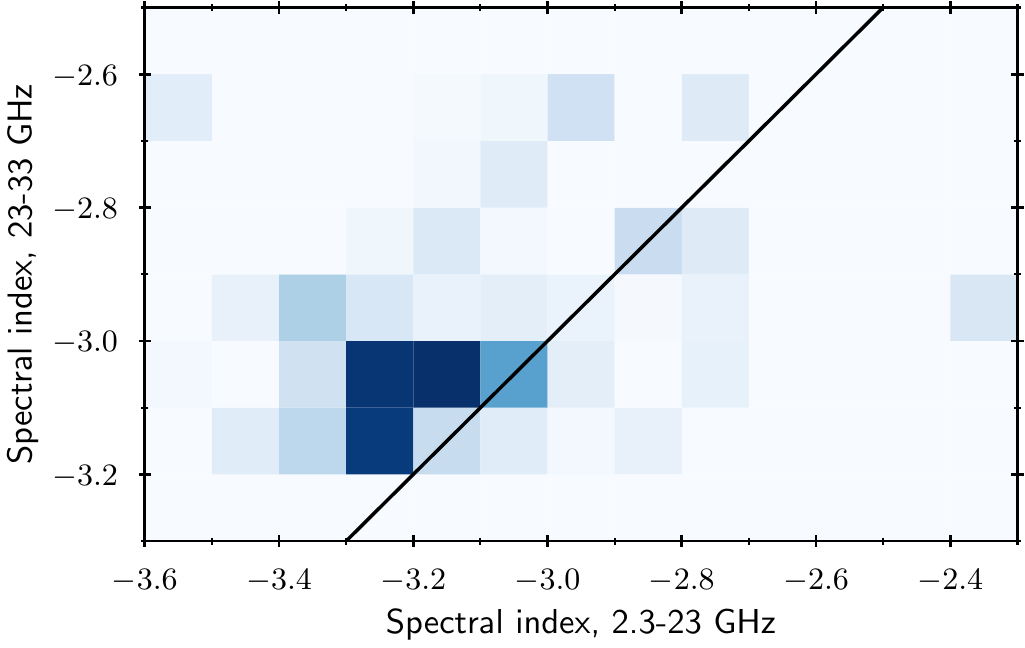}
  \end{center}
  \caption{ 2D histogram of the polarized synchrotron spectral 
    indices from  \citet{Fuskeland:2014eoa} (y-axis) versus the 
    final results in this paper (x-axis). The set of pixels that 
    are common to both studies are the final S-PASS pixels minus 
    pixels around a few bright compact objects. The line y=x is shown. 
    The figure indicates a flattening of about $\Delta\beta_s \sim 0.1$ 
    from low to higher frequencies. The histogram column with the darkest 
    blue color corresponds to about $1500$ pixels of the total $10830$ 
    pixels.}
  \label{fig:2014vs2020}
\end{figure}

Before concluding, we compare our results in the $2.3$-$23$ GHz 
frequency range to similar results in the $23$-$33$ GHz range, 
the latter obtained from \citet{Fuskeland:2014eoa} using WMAP 
K-band and WMAP Ka-band data.
The analysis performed in \citet{Fuskeland:2014eoa} is similar 
to the one in this paper, but there are two main differences. 
The first one is that the 2014 analysis reports results as a 
straight mean derived by two methods, the $T$--$T$ plot method, 
and a maximum likelihood method, while the current paper only applies 
the $T$--$T$ plot method. For more details of the maximum likelihood 
method, see \citet{Fuskeland:2014eoa}. The other difference is the 
region partitioning. Due to overall lower signal to noise ratio in 
the 2014 study, the full sky was divided into only 24 regions, where 
the diffuse regions were large and the regions inside a polarization 
foreground mask were smaller. A figure of the regions are shown in 
the top panel in Fig. 1 in \citet{Fuskeland:2014eoa}.

For this comparison study we are only interested in pixels that 
are common in both papers. We use the final S-PASS pixels, and 
results, as shown in Fig.~\ref{fig:map_betas} minus a few particular 
bright objects, including the Galactic center, which were masked out 
in the 2014 analysis. There are large uncertainties in these two 
results, due to low signal to noise ratio in the 2014 data sets, 
and the results from the current paper are prone to Faraday rotation 
mis-modeling.

Figure~\ref{fig:2014vs2020} shows a 2D histogram of the values for 
the polarized synchrotron spectral indices from \citet{Fuskeland:2014eoa} 
versus the final, Faraday rotation corrected, results in this paper. 
The straight mean of the data points from the 2014 analysis, which 
is in the $23$-$33$ GHz range, (y-axis) is $-3.01 \pm 0.14$ while 
for the data points from this paper, in the $2.3$-$23$ GHz range, 
(x-axis) is $3.15 \pm 0.21 $. The bins are of size $0.1$. Although 
the results are quite discrete because of the large regions in the 
2014 paper, it shows that there is indeed a correlation between the 
results. The correlation line of y=x is shown on the figure. The 
figure suggests a flattening of about $\Delta\beta_s \sim 0.1 \pm 0.2$ 
from low to higher frequencies. This is obtained by fitting a straight 
line with a slope equal to $1$ to the data points in the figure. The 
uncertainty is reported as the standard deviation of the residuals in 
the fit. This flattening is interesting, but in no means conclusive 
due to large uncertainties associated with these results, so we find 
no statistical evidence for curvature between $2.3$ and $33$ GHz. 
A previous analysis \citep{kogut:2012} indicates a steepening toward 
higher frequencies, by an amount of $\Delta\beta_s = -0.07$ for every 
octave in frequency. This analysis, however was done in intensity, 
and not polarization.

%Conclusion
\section{Conclusions}
\label{sec:conclusions}

In this paper we have constrained the spectral index of polarized
synchrotron emission by correlating the recently released S-PASS
2.3~GHz data with the 9-year \WMAP\ 23~GHz observations. This analysis
has been performed using a simple but robust $T$--$T$ technique,
directly correlating the two maps in pixel space, and averaging over
$15^{\circ}\times15^{\circ}$ regions. We find that the spectral index
of polarized synchrotron emission steepens from $\beta_{\mathrm{s}}\approx-2.8$ at
low Galactic latitudes to $\beta_{\mathrm{s}}\approx-3.3$ at high Galactic
latitudes, in good agreement with several previous analyses. The flat spectral 
index at the lowest Galactic latitudes are likely to be partly due to the 
depolarization effect.

A similar study based on the same data combination has already been
reported by \citet{krachmalnicoff:2019}. The main fundamental
difference between the two analyses lies in the different treatments of
Faraday rotation. 
The former analysis was made by constraining the spectral index from the polarization 
amplitude that is not affected by Faraday rotation. In addition, they masked out 
regions for which the Faraday depolarization effect was considered dominant.
In contrast, we actively correct the S-PASS observations
before correlating the linear Stokes parameters with \WMAP, and thereby 
avoid potential noise bias present in the polarization amplitude. 
Furthermore, we have considered two different
models of the rotation measure for this purpose, namely the one
presented by the S-PASS team derived directly from S-PASS, \WMAP\ and
\Planck, and one produced by \citet{Hutschenreuter:2019} based on
extra-galactic point sources and the \Planck\ free-free map. While
both models have an overall positive effect on the correlation between
S-PASS and \WMAP, the former results in a significantly higher
correlation. This is expected, given its much closer connection with
the data sets in question. Despite this important difference, we find
that our results are in good qualitative agreement with those reported
by \citet{krachmalnicoff:2019} at high Galactic latitudes. 

These results are important for future constraints on the
tensor-to-scalar ratio, as they provide insight on the overall level
of spatial variations in the synchrotron spectral index. As an
example, we applied our new constraints to estimate the level of
synchrotron emission in the BICEP2 and SPIDER regions. Overall, we
conclude the level of synchrotron emission on intermediate angular
scales in these fields are constrained to $r\lesssim0.02$ at 90~GHz
and $r\lesssim0.005$ at 150~GHz, where the upper limits are dominated
by the noise properties of \WMAP. Synchrotron emission is therefore
unlikely to pose a serious challenge for the current generation 
of $B$-mode experiments. However, if
all angular scales are considered, then the polarized synchrotron
amplitude in the BICEP2 field corresponds to a tensor-to-scalar ratio
of $r\gtrsim0.2$. This difference between large and intermediate
angular scales highlights the additional challenge required in order
to detect the $B$-mode signal at very large angular scales, as is the
target for future satellite missions such as LiteBIRD \citep{Hazumi:2019}. 
Ultimately, ancillary information from ground-based low-frequency 
experiments such as S-PASS may play a very useful role in achieving this goal.

A comparison with results obtained from \citet{Fuskeland:2014eoa} has 
allowed us to investigate the polarized synchrotron spectral index in 
two different frequency ranges. This could indicate whether the power 
law relation we have used in this paper is a pure power law, as we 
have assumed, or if it should include curvature. Our analysis suggests 
a flattening of about $\Delta\beta_s \sim 0.1 \pm 0.2$ from low to 
higher frequencies, in contrast to a previous analysis  \citep{kogut:2012} 
that indicates a steepening. This is interesting, however in no means 
conclusive due to the large uncertainties associated with this analysis, 
so we find no statistical evidence for curvature between $2.3$ and $33$ GHz, 
but we cannot rule it out either. More analyses are needed and will show 
whether the synchrotron spectral index should be modeled with or without 
a curvature.

\begin{acknowledgements}
We acknowledge support from the European Union’s Horizon 2020 research
and innovation program under grant agreement numbers 776282, 772253
and 819478, and from the Research Council of Norway. This work has made
use of S-band Polarization All Sky Survey (S-PASS) data.  Some of the
results in this paper have been derived using the HEALPix
\citep{gorski2005} software and analysis package.
\end{acknowledgements}

\bibliographystyle{aa}

%\bibliography{../common/Planck_bib,unnif_bib}
%\bibliography{Planck_bib,unnif_bib}

\begin{thebibliography}{23}
\expandafter\ifx\csname natexlab\endcsname\relax\def\natexlab#1{#1}\fi

\bibitem[{{Beck} {et~al.}(2013){Beck}, {Anderson}, {Heald}, {Horneffer},
  {Iacobelli}, {K{\"o}hler}, {Mulcahy}, {Pizzo}, {Scaife}, \&
  {Wucknitz}}]{beck:2013}
{Beck}, R., {Anderson}, J., {Heald}, G., {et~al.} 2013, Astronomische
  Nachrichten, 334, 548

\bibitem[{{Bennett} {et~al.}(2013){Bennett}, {Larson}, {Weiland}, {Jarosik},
  {Hinshaw}, {Odegard}, {Smith}, {Hill}, {Gold}, {Halpern}, {Komatsu}, {Nolta},
  {Page}, {Spergel}, {Wollack}, {Dunkley}, {Kogut}, {Limon}, {Meyer}, {Tucker},
  \& {Wright}}]{bennett2012}
{Bennett}, C.~L., {Larson}, D., {Weiland}, J.~L., {et~al.} 2013, \apjs, 208, 20

\bibitem[{{BICEP2 Collaboration} {et~al.}(2018){BICEP2 Collaboration}, {Keck
  Array Collaboration}, {Ade}, {Ahmed}, {Aikin}, {Alexand er}, {Barkats},
  {Benton}, {Bischoff}, {Bock}, {Bowens-Rubin}, {Brevik}, {Buder}, {Bullock},
  {Buza}, {Connors}, {Cornelison}, {Crill}, {Crumrine}, {Dierickx}, {Duband},
  {Dvorkin}, {Filippini}, {Fliescher}, {Grayson}, {Hall}, {Halpern},
  {Harrison}, {Hildebrand t}, {Hilton}, {Hui}, {Irwin}, {Kang}, {Karkare},
  {Karpel}, {Kaufman}, {Keating}, {Kefeli}, {Kernasovskiy}, {Kovac}, {Kuo},
  {Larsen}, {Lau}, {Leitch}, {Lueker}, {Megerian}, {Moncelsi}, {Namikawa},
  {Netterfield}, {Nguyen}, {O'Brient}, {Ogburn}, {Palladino}, {Pryke},
  {Racine}, {Richter}, {Schillaci}, {Schwarz}, {Sheehy}, {Soliman}, {St.
  Germaine}, {Staniszewski}, {Steinbach}, {Sudiwala}, {Teply}, {Thompson},
  {Tolan}, {Tucker}, {Turner}, {Umilt{\`a}}, {Vieregg}, {Wand ui}, {Weber},
  {Wiebe}, {Willmert}, {Wong}, {Wu}, {Yang}, {Yoon}, \& {Zhang}}]{bicep2}
{BICEP2 Collaboration}, {Keck Array Collaboration}, {Ade}, P.~A.~R., {et~al.}
  2018, \prl, 121, 221301

\bibitem[{Carretti {et~al.}(2019)Carretti, Haverkorn, Staveley-Smith, Bernardi,
  Gaensler, Kesteven, Poppi, Brown, Crocker, Purcell, Schnitzler, \&
  Sun}]{Carretti:2019}
Carretti, E., Haverkorn, M., Staveley-Smith, L., {et~al.} 2019, \mnras, 489,
  2330

\bibitem[{{Fixsen}(2009)}]{fixsen:2009}
{Fixsen}, D.~J. 2009, \apj, 707, 916

\bibitem[{Fuskeland {et~al.}(2014)Fuskeland, Wehus, Eriksen, \&
  N{\ae}ss}]{Fuskeland:2014eoa}
Fuskeland, U., Wehus, I.~K., Eriksen, H.~K., \& N{\ae}ss, S.~K. 2014, \apj,
  790, 104

\bibitem[{{G{\'o}rski} {et~al.}(2005){G{\'o}rski}, {Hivon}, {Banday},
  {Wandelt}, {Hansen}, {Reinecke}, \& {Bartelmann}}]{gorski2005}
{G{\'o}rski}, K.~M., {Hivon}, E., {Banday}, A.~J., {et~al.} 2005, \apj, 622,
  759

\bibitem[{Hazumi {et~al.}(2019)Hazumi, Ade, Akiba, Alonso, Arnold, Aumont,
  Baccigalupi, Barron, Basak, Beckman, Borrill, Boulanger, Bucher, Calabrese,
  Chinone, Cho, Cukierman, Curtis, de~Haan, Dobbs, Dominjon, Dotani, Duband,
  Ducout, Dunkley, Duval, Elleflot, Eriksen, Errard, Fischer, Fujino, Funaki,
  Fuskeland, Ganga, Goeckner-Wald, Grain, Halverson, Hamada, Hasebe, Hasegawa,
  Hattori, Hattori, Hayes, Hidehira, Hill, Hilton, Hubmayr, Ichiki, Iida,
  Imada, Inoue, Inoue, Irwin, Ishino, Jeong, Kanai, Kaneko, Kashima, Katayama,
  Kawasaki, Kernasovskiy, Keskitalo, Kibayashi, Kida, Kimura, Kisner, Kohri,
  Komatsu, Komatsu, Kuo, Kurinsky, Kusaka, Lazarian, Lee, Li, Linder, Maffei,
  Mangilli, Maki, Matsumura, Matsuura, Meilhan, Mima, Minami, Mitsuda, Montier,
  Nagai, Nagasaki, Nagata, Nakajima, Nakamura, Namikawa, Naruse, Nishino,
  Nitta, Noguchi, Ogawa, Oguri, Okada, Okamoto, Okamura, Otani, Patanchon,
  Pisano, Rebeiz, Remazeilles, Richards, Sakai, Sakurai, Sato, Sato, Sawada,
  Segawa, Sekimoto, Seljak, Sherwin, Shimizu, Shinozaki, Stompor, Sugai,
  Sugita, Suzuki, Suzuki, Tajima, Takada, Takaku, Takakura, Takatori, Tanabe,
  Taylor, Thompson, Thorne, Tomaru, Tomida, Tomita, Tristram, Tucker, Turin,
  Tsujimoto, Uozumi, Utsunomiya, Uzawa, Vansyngel, Wehus, Westbrook, Willer,
  Whitehorn, Yamada, Yamamoto, Yamasaki, Yamashita, \& Yoshida}]{Hazumi:2019}
Hazumi, M., Ade, P. A.~R., Akiba, Y., {et~al.} 2019, Journal of Low Temperature
  Physics, 194, 443

\bibitem[{{Hutschenreuter} \& {En\ss{}lin}(2020)}]{Hutschenreuter:2019}
{Hutschenreuter}, S. \& {En\ss{}lin}, T.~A. 2020, A\&A, 633, A150

\bibitem[{Kogut(2012)}]{kogut:2012}
Kogut, A. 2012, \apj, 753

\bibitem[{{Kogut} {et~al.}(2007){Kogut}, {Dunkley}, {Bennett}, {Dor{\'e}},
  {Gold}, {Halpern}, {Hinshaw}, {Jarosik}, {Komatsu}, {Nolta}, {Odegard},
  {Page}, {Spergel}, {Tucker}, {Weiland}, {Wollack}, \& {Wright}}]{kogut2007}
{Kogut}, A., {Dunkley}, J., {Bennett}, C.~L., {et~al.} 2007, \apj, 665, 355

\bibitem[{{Krachmalnicoff} {et~al.}(2018){Krachmalnicoff}, {Carretti},
  {Baccigalupi}, {Bernardi}, {Brown}, {Gaensler}, {Haverkorn}, {Kesteven},
  {Perrotta}, \& {Poppi}}]{krachmalnicoff:2019}
{Krachmalnicoff}, N., {Carretti}, E., {Baccigalupi}, C., {et~al.} 2018, \aap,
  618, A166

\bibitem[{{Leach} {et~al.}(2008){Leach}, {Cardoso}, {Baccigalupi}, {Barreiro},
  {Betoule}, {Bobin}, {Bonaldi}, {Delabrouille}, {de Zotti}, {Dickinson},
  {Eriksen}, {Gonz{\'a}lez-Nuevo}, {Hansen}, {Herranz}, {Le Jeune},
  {L{\'o}pez-Caniego}, {Mart{\'\i}nez-Gonz{\'a}lez}, {Massardi}, {Melin},
  {Miville-Desch{\^e}nes}, {Patanchon}, {Prunet}, {Ricciardi}, {Salerno},
  {Sanz}, {Starck}, {Stivoli}, {Stolyarov}, {Stompor}, \&
  {Vielva}}]{leach:2008}
{Leach}, S.~M., {Cardoso}, J.~F., {Baccigalupi}, C., {et~al.} 2008, \aap, 491,
  597

\bibitem[{{Liddle}(1999)}]{liddle:1999}
{Liddle}, A.~R. 1999, in High Energy Physics and Cosmology, 1998 Summer School,
  ed. A.~{Masiero}, G.~{Senjanovic}, \& A.~{Smirnov}, 260

\bibitem[{{Nagy} {et~al.}(2017){Nagy}, {Ade}, {Amiri}, {Benton}, {Bergman},
  {Bihary}, {Bock}, {Bond}, {Bryan}, {Chiang}, {Contaldi}, {Dor{\'e}},
  {Duivenvoorden}, {Eriksen}, {Farhang}, {Filippini}, {Fissel}, {Fraisse},
  {Freese}, {Galloway}, {Gambrel}, {Gandilo}, {Ganga}, {Gudmundsson},
  {Halpern}, {Hartley}, {Hasselfield}, {Hilton}, {Holmes}, {Hristov}, {Huang},
  {Irwin}, {Jones}, {Kuo}, {Kermish}, {Li}, {Mason}, {Megerian}, {Moncelsi},
  {Morford}, {Netterfield}, {Nolta}, {Padilla}, {Racine}, {Rahlin},
  {Reintsema}, {Ruhl}, {Runyan}, {Ruud}, {Shariff}, {Soler}, {Song},
  {Trangsrud}, {Tucker}, {Tucker}, {Turner}, {Van Der List}, {Weber}, {Wehus},
  {Wiebe}, \& {Young}}]{nagy:2017}
{Nagy}, J.~M., {Ade}, P.~A.~R., {Amiri}, M., {et~al.} 2017, \apj, 844, 151

\bibitem[{Orear(1982)}]{orear1982}
Orear, J. 1982, American Journal of Physics, 50, 912

\bibitem[{{Page} {et~al.}(2003){Page}, {Barnes}, {Hinshaw}, {Spergel},
  {Weiland}, {Wollack}, {Bennett}, {Halpern}, {Jarosik}, {Kogut}, {Limon},
  {Meyer}, {Tucker}, \& {Wright}}]{page2003a}
{Page}, L., {Barnes}, C., {Hinshaw}, G., {et~al.} 2003, \apjs, 148, 39

\bibitem[{{Planck Collaboration} {et~al.}(2015){Planck Collaboration}, {Ade},
  {Aghanim}, {Alina}, {Alves}, {Armitage-Caplan}, {Arnaud}, {Arzoumanian},
  {Ashdown}, {Atrio-Barand ela}, {Aumont}, {Baccigalupi}, {Banday}, {Barreiro},
  {Battaner}, {Benabed}, {Benoit-L{\'e}vy}, {Bernard}, {Bersanelli},
  {Bielewicz}, {Bock}, {Bond}, {Borrill}, {Bouchet}, {Boulanger}, {Bracco},
  {Burigana}, {Butler}, {Cardoso}, {Catalano}, {Chamballu}, {Chary}, {Chiang},
  {Christensen}, {Colombi}, {Colombo}, {Combet}, {Couchot}, {Coulais}, {Crill},
  {Curto}, {Cuttaia}, {Danese}, {Davies}, {Davis}, {de Bernardis}, {de Gouveia
  Dal Pino}, {de Rosa}, {de Zotti}, {Delabrouille}, {D{\'e}sert}, {Dickinson},
  {Diego}, {Donzelli}, {Dor{\'e}}, {Douspis}, {Dunkley}, {Dupac}, {Efstathiou},
  {En{\ss}lin}, {Eriksen}, {Falgarone}, {Ferri{\`e}re}, {Finelli}, {Forni},
  {Frailis}, {Fraisse}, {Franceschi}, {Galeotta}, {Ganga}, {Ghosh}, {Giard},
  {Giraud-H{\'e}raud}, {Gonz{\'a}lez-Nuevo}, {G{\'o}rski}, {Gregorio},
  {Gruppuso}, {Guillet}, {Hansen}, {Harrison}, {Helou},
  {Hern{\'a}ndez-Monteagudo}, {Hildebrand t}, {Hivon}, {Hobson}, {Holmes},
  {Hornstrup}, {Huffenberger}, {Jaffe}, {Jaffe}, {Jones}, {Juvela},
  {Keih{\"a}nen}, {Keskitalo}, {Kisner}, {Kneissl}, {Knoche}, {Kunz},
  {Kurki-Suonio}, {Lagache}, {L{\"a}hteenm{\"a}ki}, {Lamarre}, {Lasenby},
  {Lawrence}, {Leahy}, {Leonardi}, {Levrier}, {Liguori}, {Lilje},
  {Linden-V{\o}rnle}, {L{\'o}pez-Caniego}, {Lubin}, {Mac{\'\i}as-P{\'e}rez},
  {Maffei}, {Magalh{\~a}es}, {Maino}, {Mandolesi}, {Maris}, {Marshall},
  {Martin}, {Mart{\'\i}nez-Gonz{\'a}lez}, {Masi}, {Matarrese}, {Mazzotta},
  {Melchiorri}, {Mendes}, {Mennella}, {Migliaccio}, {Miville-Desch{\^e}nes},
  {Moneti}, {Montier}, {Morgante}, {Mortlock}, {Munshi}, {Murphy}, {Naselsky},
  {Nati}, {Natoli}, {Netterfield}, {Noviello}, {Novikov}, {Novikov},
  {Oxborrow}, {Pagano}, {Pajot}, {Paladini}, {Paoletti}, {Pasian}, {Pearson},
  {Perdereau}, {Perotto}, {Perrotta}, {Piacentini}, {Piat}, {Pietrobon},
  {Plaszczynski}, {Poidevin}, {Pointecouteau}, {Polenta}, {Popa}, {Pratt},
  {Prunet}, {Puget}, {Rachen}, {Reach}, {Rebolo}, {Reinecke}, {Remazeilles},
  {Renault}, {Ricciardi}, {Riller}, {Ristorcelli}, {Rocha}, {Rosset},
  {Roudier}, {Rubi{\~n}o-Mart{\'\i}n}, {Rusholme}, {Sandri}, {Savini}, {Scott},
  {Spencer}, {Stolyarov}, {Stompor}, {Sudiwala}, {Sutton}, {Suur-Uski},
  {Sygnet}, {Tauber}, {Terenzi}, {Toffolatti}, {Tomasi}, {Tristram}, {Tucci},
  {Umana}, {Valenziano}, {Valiviita}, {Van Tent}, {Vielva}, {Villa}, {Wade},
  {Wandelt}, {Zacchei}, \& {Zonca}}]{planck_PIP_XIX}
{Planck Collaboration}, {Ade}, P.~A.~R., {Aghanim}, N., {et~al.} 2015, \aap,
  576, A104

\bibitem[{{\sorthelp{Planck Collaboration 2018A}}{Planck Collaboration
  I}(2020)}]{planck2016-l01}
{\sorthelp{Planck Collaboration 2018A}}{Planck Collaboration I}. 2020, \aap,
  641, A1

\bibitem[{{\sorthelp{Planck Collaboration 2018B}}{Planck Collaboration
  II}(2020)}]{planck2016-l02}
{\sorthelp{Planck Collaboration 2018B}}{Planck Collaboration II}. 2020, \aap,
  641, A2

\bibitem[{{\sorthelp{Planck Collaboration 2018D}}{Planck Collaboration
  IV}(2020)}]{planck2016-l04}
{\sorthelp{Planck Collaboration 2018D}}{Planck Collaboration IV}. 2020, \aap,
  641, A4

\bibitem[{{\sorthelp{Planck Collaboration 2018F}}{Planck Collaboration
  VI}(2020)}]{planck2016-l06}
{\sorthelp{Planck Collaboration 2018F}}{Planck Collaboration VI}. 2020, \aap,
  641, A6

\bibitem[{{Zaldarriaga} \& {Seljak}(1997)}]{zaldarriaga:1997}
{Zaldarriaga}, M. \& {Seljak}, U. 1997, \prd, 55, 1830

\end{thebibliography}

%\appendix

%\section{Appendix}
%\label{app:appendix}

\end{document}